\documentclass[twocolumn,preprintnumbers,amsmath,amssymb,superscriptaddress]{revtex4-2}
\usepackage{graphicx}
\usepackage[colorlinks=true,urlcolor=blue,citecolor=blue,linkcolor=blue]{hyperref}

\usepackage{amsmath}
\usepackage{amssymb}
\usepackage{amsthm}
\usepackage{amsfonts}
\usepackage{enumerate}

\usepackage{centernot}
\usepackage{tikz}

\usepackage{subfigure}
\usepackage{ulem}
\definecolor{LL-color}{named}{blue}
\newcommand{\LLcmt}[1]{\textcolor[named]{LL-color}{{\bf LL: #1 }}}

\newcommand{\affiliationSICNU}{Department of Physics, Institute of Solid State Physics and Center for Computational Sciences, Sichuan Normal University, Chengdu, Sichuan 610066, China}

\newcommand{\affiliationFDU}{Department of Physics, Fudan University, Shanghai 200433, China}

\newcommand{\affiliationUSST}{College of Science, University of Shanghai for Science and Technology, Shanghai 200093, China}

\begin{document}

	\title{Non-Fermi Liquid Behavior of the $t$-$J$ Model in the Strange Metal Phase: $U(1)$ Gauge Theory Consistent with Local Constraints}
	\author{Long Liang}
	\affiliation{\affiliationSICNU}

	\author{ Yue Yu}
	\thanks{Correspondence to: yuyue@fudan.edu.cn}
	\affiliation{\affiliationFDU}
	\author{Xi Luo}
	\thanks{Correspondence to: xiluo@usst.edu.cn}
	\affiliation {\affiliationUSST}

	\date{\today}
	
	\begin{abstract}
	    {In the slave particle representation with $U(1)$ gauge symmetry, local constraints on physical states characterized by various mean field solutions belong to Dirac's second-class ones.
        Although constrained systems are extensively investigated}, realistic methods to solve the gauge theory problem with second-class constraints are yet to be developed. 
        We formulate a Becchi-Rouet-Stora-Tyutin (BRST) quantization theory,  {called consistent $U(1)$ gauge theory}, that is consistent with  both first- and second-class local constraints for strongly correlated condensed matter systems. 
         {In our consistent $U(1)$ gauge theory}, the redundant gauge degrees of freedom are removed by proper gauge fixing conditions while the constraints are exactly retained and  the gauge invariance is guaranteed by the BRST symmetry. 
        Furthermore, the gauge fixing conditions endow the gauge field with dynamics. 
        This turns the strongly correlated electron model into a weakly coupled slave boson model,  {so} most of the system's physical  {properties} can be calculated by the conventional quantum many-body perturbation method. 
        We focus on the property of the strange metal phase in the $t$-$J$ model. 
        The electron momentum distribution and the spectral function are calculated, and the non-Fermi liquid behavior agrees with the angle-resolved photoemission spectroscopy measurements for cuprate materials. 
        We also study the electromagnetic responses of the strange metal state. 
        The observed non-Fermi liquid anomalies are captured by our calculations. 
        Especially, we find that the Hall resistivity decreases as temperature increases, and the sign of the Hall resistivity varies from negative to positive when the dopant concentration varies from optimal doping  to underdoping  {in the strange metal regime.}
	\end{abstract}
	
	\maketitle
	
	\section{Introduction}
	
	 {The gauge principle plays a fundamental role in our understanding of various  phenomena in diverse physical systems, ranging from high energy to condensed matter physics.} 
 The slave boson/fermion  representation of the electron operator~\cite{Bar,Bar2,Col,KL,SHF,LWH,Ru,zou,affl,AA, Yosh,Cha} has an intrinsic gauge symmetry and is a powerful tool in the study of strongly correlated condensed matter systems~\cite{NLPRL,LN,WL,WNL,Lee,MuF,MuF2,Sach,Sach1,Bonetti}.
	
	The constraints on the local quantum states are serious obstacles  {to} solving strongly correlated problems.  A well-known example is no double occupation of the electron at one lattice site in the $t$-$J$ model~\cite{tJ1,tJ2,ZR} where the Gutzwiller projected variational  wave functions (or renormalized mean field theory) are usually used~\cite{Gutzwiller_1963,B,RMFT,RGU1,RGU2,RGU3}.  The statistically-consistent Gutzwiller projection~\cite{SGA,SGA1,SGA2,SGA3,SGA4} is equivalent to the slave boson mean-field theory.  
	%
	%
	Based on the Faddeev-Jackiw approach~\cite{FJ} to constrained systems,  an $X$-operator formalism  {that} can be mapped to the slave particle representation was developed~\cite{X-O} and was used to calculate the spectral function of the  {pseudogap} phase~\cite{Greco}. 
	
	Besides various analytical methods, large-scale numerical simulation techniques, such as quantum Monte Carlo~\cite{QMC,QMC1}, functional~\cite{FRG}, density matrix~\cite{DMRG}, tensor network~\cite{TN} renormalization group methods, as well as dynamical mean field theory~\cite{DMFT,Singh},  have been greatly developed to solve strongly correlated problems.  
 It is far beyond the scope of this article to summarize the analytical and numerical developments. 
	
The Hubbard model and the  $t$-$J$ model were suggested  {as} the simplest models to explain the basic physics of anomalous properties of cuprates~\cite{ZR,Anderson}, which are strongly correlated materials exhibiting high-$T_c$ superconductivity~\cite{highTCSC}. {The recently discovered high-$T_c$  nickelate superconductor under pressure La$_3$Ni$_2$O$_7$ \cite{ni-exp} is also believed to be described by a bilayer $t$-$J$ model~\cite{ni1,ni2,ni3,ni4,ni5,ni6,ni7,ni8}.} The slave boson mean field theory may qualitatively capture the phase diagram for cuprate and nickelate materials. Recently, the Rashba-type spin-orbital coupling on the surface of Bi$_2$Sr$_2$CaCu$_2$O$_{8+d}$  {was} reported~\cite{ashvin2018}, and the corresponding theoretical proposals on possible nonlinear and nonreciprocal transport phenomena, which are based on the slave boson theory with the Rashba spin-orbital coupling, may provide a new angle of view to explore the phases and electronic states in high-$T_c$~cuprates~\cite{nagaosa2023}. The gauge theory has been developed in~\cite{NLPRL,LN,WL,WNL,Lee} to obtain more quantitative results  {that} can be compared to the experimental data for cuprates. However, since the gauge field in that theory is not dynamical in the first place, only the electromagnetic responses can be  {calculated} microscopically under the Gaussian approximation after integrating over the holon and spinon fields. The electronic  properties such as the spectral function,  self-energy, and pairing gap can only be calculated by either introducing phenomenological approximations or at the mean field level. 
	
In a recent work, we found that the previous studies of the gauge theory based on the slave boson/fermion representation  {are} incomplete and further gauge fixing conditions are needed~\cite{LLY}.  For example, in the slave boson representation, the no-double occupancy constraint for the $t$-$J$ model reduces to the condition that only one holon or one spinon can occupy a local lattice site. Relaxing this constraint  breaks the fermionic nature of the electron and introduces unphysical degrees of freedom. Therefore, the saddle point approximation of the slave boson mean field theory is unreliable and uncontrollable if the gauge fluctuations around the saddle point do not restore the local constraint. We showed that to retain the local constraint, proper gauge fixing conditions have to be added to remove the redundant gauge degrees of freedom in the Lagrange multiplier that enforces the local constraint. Some features of Dirac's  {first-class} constraint~\cite{Dirac1,Sundermeyer} for the $t$-$J$ model in the slave particle representation have been studied~\cite{GuRa}. In fact, the procedure to introduce the gauge fixing condition is equivalent to the gauge fixing procedure  {for} Dirac's first-class constrained systems developed by Fradkin, Vilkovisky, and Batalin (FVB)~\cite{FV1,FV2,BV}. The presence of the Becchi-Rouet-Stora-Tyutin (BRST) symmetry~\cite{BRS,Ty,Ty1} after gauge fixing is the criterion  {for determining} whether the gauge fixing condition is consistent with the local constraint or not.  {This is} because the requirement of BRST invariance is exactly equivalent to Dirac's first-class constraints. 

In the trivial atomic limit of the $t$-$J$ model, the local constraints are the first-class ones; however, as we will see, the constraints  {on} all other ordered mean field states are Dirac's second-class ones as the local constraints of vanishing counterflow between the holon and spinon currents are enforced.  The FVB's BRST procedure~\cite{FV1,FV2,BV} does not work for the second-class constraints in these ordered mean field states in which violations of the Fermi liquid behavior  are discovered.  As recognized in the previous gauge theory~\cite{IL,NLPRL,LN}, the spatial components of the gauge field should be introduced to recover the gauge invariance. 

In this work, we  develop a BRST quantization procedure for a realistic physical system with  second-class constraints for the first time. We then provide a reliable and effective method for dealing with the stability of the ordered mean field states of a strongly correlated system under quantum fluctuations with a controlled perturbation calculation.

The problems in the previous gauge theory are, in turn, that (a)  {since} the spatial components of the gauge field play the role of the Lagrange multipliers that enforce the vanishing counterflow between the holon and spinon currents,   {a} gauge fixing condition consistent with the vanishing counterflow constraint is also required. But this was not considered in the previous theory.  (b)  The coupling constant of the previous gauge field theory is in the strong coupling limit, and there is no small parameter for perturbation calculations. (c) To obtain the dynamics of the gauge field in the previous method, one must integrate over the spinon and holon, but this leads to the problem that many physical  {observables} cannot be directly calculated.
	
By using our BRST quantization procedure, we show that the  {aforementioned} problems can be solved. The gauge field acquires  dynamics due to the consistent choice of the gauge fixing conditions, and the coupling constant of the gauge theory becomes finite. 
For the $U(1)$ gauge theory, the problem can  be solved without using BRST quantization in general. However, in such a constrained system, it is not easy to check if the gauge fixing conditions are consistent with the constraints.  With the BRST quantization, this consistency can be easily checked according to the BRST symmetry, which requires that the physical states are BRST charge-free. {However, as we mentioned, a new BRST procedure beyond the FVB theory needs to be developed for the mean field states  {because}  both  the local constraints of no-double occupation and the vanishing counterflow  are no longer the Dirac's first-class ones but are  second-class ones.}  Although several formal developments in applying BRST methods to systems with second-class constraints were proposed~\cite{second1,second2,second3,second4,second5}, we find that these methods cannot be  applied to our condensed matter systems. Fortunately, for the $t$-$J$ model, we find that the BRST symmetry consistent with these second-class constraints exists. The BRST  {is  charge-free in} the physical states, and the Euler-Lagrange equations self-consistently recover the original constraints and the gauge fixing conditions.   In this way, we obtain a well-defined perturbation theory with a weak coupling constant. 

In the present work, we focus on the properties  {of} the strange metal phase of the $t$-$J$ model and leave those of the Fermi liquid, pseudogap, superconducting, and  antiferromagnetic phases for further study. In the weak coupling region, the mean field state of the strange metal phase is the uniform resonant valence bond (uRVB) state proposed by Anderson~\cite{Anderson}. We calculate the electron momentum distribution and the electron spectral function in the strange metal phase. The non-Fermi liquid behavior with the spin-charge separation is explicitly shown through the  {zero-temperature} electron momentum distribution, which  {violates} Luttinger's theorem~\cite{Lutt}. The electron spectral function  {in the strange metal phase} coincides with the angle-resolved photoemission spectroscopy (ARPES) measurement data for cuprates~\cite{ARPES1,ARPES2,ARPES}. Our results substantially improve  {upon} previous  {findings} by Anderson and Zou~\cite{AZ},  {as well as} the gauge theory~\cite {NLPRL,LN}. The electromagnetic responses are also  {calculated} perturbatively. We see that there is a temperature region where the resistivity depends linearly on temperature, which is an anomalous phenomenon in a wide temperature region for the  {optimally} doped cuprates~\cite{Kuz}.  As an improvement of the approximate analytical calculation in~\cite{NLPRL,LN}, we numerically calculate the Hall resistivity  and find that in the strange metal phase,  the Hall resistivity decreases as temperature  {increases}, and the sign of the Hall resistivity changes from negative to positive when the dopant concentration $x$ varies from the optimal doping one to the underdoping side. Our result is consistent with the behavior of the Hall resistivity observed experimentally~\cite{Fu,Hall1,Hall2},  {and in our theory, additional scattering time~\cite{chien} is not needed.}

This paper is organized as follows: In Sec.~\ref{sectionII},  we quantize the $t$-$J$ model by using the path integral within the BRST formalism. In Sec.~\ref{sectionIII}, we  {focus on the strange metal phase and present the noninteracting gauge propagators and interaction vertices.} 
In Sec.~\ref{sectionIV}, we calculate the momentum distribution of the electrons in the strange metal phase and show the non-Fermi liquid nature. We also calculate the electron spectral function for the strange metal phase and compare the result to the APRES data. 
 {Possible violation of  Luttinger's theorem is discussed.} In Sec.~\ref{sectionV}, we study the electromagnetic responses to the external  electric and magnetic fields. The longitudinal resistivity and the Hall resistivity are calculated. Especially, the dependence of the Hall resistivity on temperature and the dopant concentration are presented.  Sec.~\ref{sectionVI} is devoted to conclusions and perspectives. 
	
\section{consistent $U(1)$ gauge Theory for the $t-J$ model} \label{sectionII}


The Hamiltonian for the $t$-$J$ model on a square lattice is given by
\begin{eqnarray}
  H_{t-J}=-t\sum_{\langle ij\rangle,\sigma}c^\dag_{i\sigma}c_{j\sigma}+J\sum_{\langle ij\rangle}({\bf S}_i\cdot{\bf S}_j-\frac{1}4 n_in_j), \label{tJH}
\end{eqnarray}
where $c_{i\sigma}$ is the electron annihilation operator at a lattice site $i$ with spin $\sigma$;  $S^a_i=\frac{1}2 \sum_{\sigma,\sigma'}c^\dag_{i\sigma}\sigma^a_{\sigma\sigma'}c_{i\sigma'}$ are the spin operators, and $\sigma^a$ ($a=x,y,z$) are Pauli matrices.  The hopping amplitude $t$ and the exchange amplitude $J$ are fixed in between the nearest neighbor sites.  The constraint is that there is no double occupation at each lattice site, i.e., $c^\dag_ic_i\leq 1$ for all $i$ with a fixed total electron number.  
	
In the slave boson representation,  the electron operator is decomposed into the fermionic spinon and bosonic holon, $c^\dag_{i\sigma}= f^\dag_{i\sigma}h_i$,  where $f^\dag_{i\sigma}$ is the spinon creation operator and $h_i$ is the holon annihilation operator.  This decomposition works when the local constraint  {is enforced for every site $i$ by}
\begin{eqnarray}
	G_i=h_i^\dag h_i+\sum_\sigma f_{i\sigma}^\dag f_{i\sigma}-1=0.
\end{eqnarray}
The electron number is fixed so that the average density of the spinon is $1-x$, where $x$ is the holon concentration.  Notice that if we write  the Hamiltonian~\eqref{tJH} in the slave boson representation, it involves only the operator combinations  $f^\dag_{i\sigma }f_{i\sigma' }$, $f_{i\sigma }h^\dag_i$, {\it etc.} at any local site $i$. Thus, one has $[G_i,H_{t-J}]=0$, and $G_i=0$ is the first-class constraint.

 {In Lagrangian field theory, the local constraints can be imposed by introducing Lagrange multipliers for every lattice site.} For finite temperature and in the imaginary time formalism, the effective slave boson $t$-$J$ Lagrangian reads~\cite{NLPRL,LN} 
 {
	\begin{eqnarray}\label{t-J}
 L_\lambda&=&\sum_i h^\dag_i\partial_\tau h_i+\sum_{i\sigma}f^\dag_{i\sigma}(\partial_\tau-\mu) f_{i\sigma}\nonumber\\
 &&-ig\sum_i\lambda_iG_i+H_{t-J},
	\end{eqnarray}}
where $0<\tau<\beta$ with $\beta=1/T$ being the inverse of temperature. 
  {Note that the Lagrange multipliers $\lambda_i$ are treated as dynamical variables, which become time-dependent and can be identified as the temporal component of the $U(1)$ gauge potential. This introduces redundant degrees of freedom that should be eliminated through a suitable gauge fixing condition that is consistent with local constraints.}
 
 The Lagrangian Eq.~\eqref{t-J} is gauge invariant under the transformation $(f_{i\sigma}(\tau),h_i(\tau))\to e^{ig\theta_i(\tau)}(f_{i\sigma}(\tau),h_i(\tau))$, and $\lambda_i(\tau)\to \lambda_i(\tau)+\partial_\tau\theta_i(\tau)$. The  {gauge-invariant} partition function is
	\begin{eqnarray}
		Z=\int \prod_{i,\tau} d\Phi_i(\tau)e^{-\int_0^\beta d\tau L_\lambda},
	\end{eqnarray} 
	where $\Phi_i(\tau)$  {represents} for all fields, $\lambda_i(\tau)$, $h_i(\tau)$,$h^\dag_i(\tau)$, $...$, and $g$ is introduced as an arbitrary coupling constant because $$\int d\lambda_i(\tau) e^{ig\lambda_i(\tau)G_i}=\delta(gG_i)$$ also  {imposes} the constraint $G_i=0$.
	As we have done in \cite{LLY}, for such a constrained gauge  theory, in order to remove the redundant gauge degrees of freedom while keeping  the partition function gauge invariant, a proper gauge fixing condition for $\lambda_i(\tau)$ is needed.  
	
	\subsection{The BRST quantization for the exact theory}\label{subsection_BRST_exact}
	
	Since $G_i=0$ is the first-class constraint, one can apply FVB's procedure to quantize the theory \cite{FV1,FV2,BV}. 
	In our previous work \cite{LLY}, we  {used} the gauge fixing condition 
	\begin{eqnarray}
\partial_\tau\lambda_i(\tau)=\xi_1\pi_{\lambda_i}, \label{lambdaG1}
	\end{eqnarray}
	where $\xi_1$ is an arbitrary constant and $\pi_{\lambda_i}(\tau)$  is the canonical conjugate of $\lambda_i(\tau)$ \cite{FV1,FV2,BV}.  {For a detailed explanation {of} why this gauge fixing is necessary, see Appendix~\ref{ABRST}.}  Physically, since there is no `potential energy' of $\lambda_i$, it cannot be `accelerated'.  The Euler-Lagrange equation of $\lambda_i$ confirms this point, i.e.,
	\begin{eqnarray}
		\partial^2_\tau\lambda_i=\partial_\tau \pi_{\lambda i}=0. \label{dotpi}
	\end{eqnarray}
	Applying to the $t$-$J$ model, the BRST invariant Lagrangian, which removes the redundant gauge degrees of freedom, is given by  
	\begin{eqnarray}
		L^{(1)}_{BRST}=L_\lambda-\frac{1}{2\xi_1}\sum_i(\partial_\tau\lambda_i)^2 +\sum_i\bar u_{1i}\partial^2_\tau u_{1i},
	\end{eqnarray}
	where $u_{1i}(\tau)$ and $\bar u_{1i}(\tau)$ are the ghost and anti-ghost fields satisfying  {the} anticommutation relation $\{u_{1i},\bar u_{1j}\}=\delta_{ij}$; and the BRST transformations read $\delta_{B1}f_\sigma=i\epsilon gu_1f_\sigma, \delta_{B1}h=i\epsilon gu_1h,\delta_{B1} \lambda=\epsilon \partial_\tau u_1,\delta_{B1}u_1=0,$ and $\delta_{B1}\bar u_1=\epsilon\partial_\tau\lambda/\xi_1$, where $\epsilon$ is a Grassmann constant with $\epsilon^2=0$. 
	
	The BRST invariant partition function reads
	\begin{eqnarray}
		Z^{(1)}_{BRST}=\int \prod_{i,\tau}d\Phi_i(\tau)d\bar u_{1i}(\tau)du_{1i}(\tau)e^{-\int_0^\beta d\tau L^{(1)}_{BRST}}.
	\end{eqnarray}
 Up to a constant factor, this partition function is  {gauge-invariant}.
	The Euclidean BRST charge is given by 
	\begin{eqnarray}
		Q_{B1}=\sum_i(igG_i u_i-\frac{1}{\xi_1}(\partial_\tau\lambda_i)\partial_\tau u_{1i}).
	\end{eqnarray}
	 {One can check $[Q_{B1},H^{(1)}_{BRST}]=0$ with $H^{(1)}_{BRST}$ being the corresponding Hamiltonian, which implies the BRST invariance of the theory. }
	Since the BRST symmetry is a global symmetry, the physical states of the theory are the eigenstates of the BRST charge. Furthermore, the gauge fixing terms together with the ghost part in a general BRST invariant theory can be rewritten in  {the} form of the BRST transformation of a functional with the ghost number $-1$, i.e., a BRST exact form~\cite{Weinberg}.  Therefore, the variation of any matrix element between two physical states $\langle {\rm phys}'|{\rm phys}\rangle$ in an arbitrary BRST transformation under which the action is invariant vanishes identically. Consequently,  any physical state must be BRST  {charge-free}~\cite{Weinberg}. 
	For this model, we can easily  {verify} this result, and then  the physical states satisfy
	\begin{eqnarray}
		Q_{B1}|{\rm phys}\rangle=0.
	\end{eqnarray}
	Furthermore, since $u_i(\tau)$ and $\partial_\tau u_i(\tau)$ are independent local fields, the constraints $G_i=0$ and $\partial_\tau\lambda_i=0$ are exactly recovered. {\it This means that the BRST quantization consistently combines the local constraints and the gauge fixing conditions in a  {systematic} way}.    
	
	Notice that for the Abelian gauge theory, the ghost sector is decoupled from the gauge field. One can integrate over the ghost sector, and this leads to a determinant  {that} is independent of the `matter' sector. Dropping this constant determinant, we arrive at the gauge fixed partition function,  where the gauge symmetry seems to be broken but
	actually not because  {the gauge symmetry is now replaced by the global BRST symmetry and the locality in the gauge transformation is hidden in the ghost field.}
 Thus, for the $U(1)$ gauge theory, the effective partition function can be written as
	\begin{eqnarray}
		Z^{(1)}_{eff}=\int \prod_{i,\tau}d\Phi_i(\tau) e^{-\int_0^\beta d\tau L^{(1)}_{eff}}, \label{ET}
	\end{eqnarray}
	where 
	\begin{eqnarray}
		L^{(1)}_{eff}=L_\lambda-\frac{1}{2\xi_1}\sum_i(\partial_\tau\lambda_i)^2.\label{ETL}
	\end{eqnarray}
	So far, the theory for $L^{(1)}_{eff}$ is exactly equivalent to the original $t$-$J$ model. Although the BRST formalism is not actually used in the effective theory, it ensures consistency between the gauge fixing condition and the local constraint. Furthermore, it is easy to check that the BRST operator is nilpotent $Q_{B1}^2=0$, and the BRST cohomology could be used to classify the topology of the physical state space. 
	
 For a non-Abelian gauge theory, the ghost part cannot be dropped because, in that case, the ghost part of the Lagrangian does depend on the gauge field, i.e., the ghost sector couples with the gauge field in general.  
 The BRST formalism will greatly simplify the consistent quantization of the $SU(2)$ gauge theory of the slave boson~\cite{WNL,WWNL}, which goes beyond the scope of this work and will not be discussed further.
	
	\subsection{The BRST quantization for ordered gauge theory}
	
The theory described by Eq.~\eqref{ET} is exactly equivalent to the original $t$-$J$ model. However, if we perform perturbation calculations around a trivial non-ordered ground state, we cannot arrive at  ordered states, which may be more favorable in energy below some characteristic temperature. A common strategy to obtain these ordered phases is to do mean field  {approximations}. Previously, we used the BCS  mean field theory to study the superconducting state~\cite{LLY}. We show that after integrating out  {the} $\lambda_i(\tau)$ field, the second term of Eq.~\eqref{ETL}  gives an extra dynamic pairing term, which corrects the conventional mean field pairing term. As a result, the mean field pairing gap is suppressed.  In this section, we would like to develop a complete gauge theory based on various mean field states  {using} BRST quantization. To this end,  we first write the $U(1)$ gauge theory in the slave boson mean field approximation~\cite{NLPRL,LN}, 
	\begin{eqnarray}
		L_{MF}&=&\frac{J}4 \sum_{\langle ij\rangle}[|\gamma^f|^2+|\Delta_a|^2-\sum_\sigma (\gamma^{f\dag}e^{i a_{ij}}f^\dag_{i\sigma} f_{j\sigma}+h.c.)]\nonumber\\
		&+&\sum_{\langle ij\rangle}\frac{J}4[\Delta_ae^{i\phi_{ij}}(f_{i\uparrow}^\dag f_{j\downarrow}^\dag-f_{i\downarrow}^\dag f_{j\uparrow}^\dag)+h.c.]\nonumber\\
		&+&\sum_ih^\dag_i(\partial_\tau-\mu_h) h_i+\sum_{i\sigma}f^\dag_{i\sigma}(\partial_\tau -\mu_f)f_{i\sigma}\nonumber\\
		&-&t\sum_{\langle ij\rangle}(e^{ia_{ij}}(\gamma^f h^\dag_i h_j+\gamma^{h\dag} f^\dag_{i\sigma} f_{j\sigma})+h.c.)\nonumber\\
		&+&\sum_i ig\lambda_iG_i, \label{LMF}
	\end{eqnarray}
	where $\Delta_a$ for $a=x,y$ labels the pairing parameter in the $a$-link, and $\gamma ^{h,f}$ are the hopping parameters for the spinon and the holon.   {We} choose $\Delta_a$ and $\gamma ^{h,f}$  {as} expectation values in the mean field approximation. The phase fields $a_{ij}$ and $\phi_{ij}$, which obey the periodic boundary condition, are  quantum fluctuations to compensate  {for} the gauge symmetry breaking due to the mean field approximation.  Besides the temporal gauge field  $\lambda_i$,  the mean field theory has a spatial gauge invariance under the transformations $(f_{i\sigma}(\tau),h_i(\tau))\to e^{i\theta_i(\tau)}(f_{i\sigma}(\tau),h_i(\tau))$, and $a_{ij}\to a_{ij}+\theta_i-\theta_j$ and $\phi_{ij}\to  \phi_{ij}+\theta_i+\theta_j$. 
  {The equation of motion of $a_{ij}$ leads to the constraint of vanishing counterflow between the holon and spinon currents}~\cite{NLPRL,LN}, i.e.,
	\begin{eqnarray}
		J_{ij}=J^f_{ij}+J^h_{ij}=0.\label{cconstraint}
	\end{eqnarray} 
	This is also a local constraint.  {Note that the gauge field appears  in the expression of the spinon and holon currents. And the constraint holds for non-vanishing gauge configurations.} The variation of $\phi_{ij}$  does not result in new constraints, and this point will be further analyzed in Sec.~\ref{SectionIIC}.  Therefore, our problem becomes how to quantize the mean field theory with the constraints $G_i=0$ and $J_{ij}=0$ with proper gauge fixing conditions. 

 	At the mean field approximation, one can take  $a_{ij}=\bar a_{ij}$,  {with $\bar{a}_{ij}$ being a background gauge configuration determined by solving the mean field self-consistent equations.} To  {maintain} translation symmetry as well as time reversal symmetry, the gauge flux through a plaquette takes $0~{\rm or}~\pi~{\rm mod}(2\pi)$, which corresponds to either the uRVB state~\cite{Anderson} or the $\pi$-flux state~\cite{affl}.
	The mean field energy of the former is lower than that of  {the} latter in the strange metal phase \cite{GK}. Following Nagaosa and Lee \cite{NLPRL,LN}, we take the uRVB mean field state. However, we still need to deal with the gauge fluctuation $g\delta a_{ij}=a_{ij}-\bar a_{ij}$  around the uRVB state. 
	
	
	In the continuum limit, $G_i$ changes to $G({\bf r})$, and $\bar a_{ij}$ changes to $\bar{\mathbf{a}}(\mathbf{r})$, 
 which is zero in the uRVB state, and $\phi({\bf r},\tau)$ is the continuum limit of $(\phi_i+\phi_j)/2$. The conserved current $J_\mu$ is given by $J_\tau=G+1$ and
	\begin{eqnarray}
		&& J_b(\delta a)= J_{fb}(\delta a)+J_{hb}(\delta a)\nonumber\\
		&&=-\frac{1}{m_f}\sum_\sigma f_\sigma^\dag(i\partial_b+g \delta a_b)f_\sigma
		-\frac{1}{m_h}h^\dag(i\partial_b+g \delta a_b)h,\nonumber
	\end{eqnarray}
	where $1/m_h\sim \gamma^ft$, $1/m_f\sim \gamma^f J+\gamma^h t$, and $\delta a_b$ is the continuum limit of $\delta a_{ij}$, namely, $\delta a_{ij}=({\bf r}_i-{\bf r}_j)\cdot \mathbf{a}[({\bf r}_i+{\bf r}_j)/2]$. We also set $g\lambda({\bf r},\tau)=g \delta \lambda$.
	The constraint \eqref{cconstraint} becomes $J_b(\delta a)=0$,  {which comes from the equation of motion of the spatial gauge potential $\delta a_b$}. This current  $J_b(\delta a)$ is gauge invariant but not a physical observable  because it is dependent on the gauge fluctuation. The physical  observable vanishing counterflow constraint is given by
	\begin{eqnarray}
		J_b=\langle J_b(\delta a)\rangle_{\delta a}=0,
	\end{eqnarray} 
	where $\langle\cdots \rangle_{\delta a}$ stands for integrating the gauge fluctuation $\delta a_b$. However, due to the redundant gauge degrees of freedom in integrating over $\delta a_b$, we need to choose the gauge fixing conditions  {that} are consistent with our constraints. 
	
	Before proceeding with quantization, we follow Dirac's classification~\cite{Dirac1} and check the class of the constraints.  It is easy to read out the mean field Hamiltonian from the Lagrangian~\eqref{LMF} and  we find that in the continuum limit  
	\begin{eqnarray}
		[H_{MF}, G({\bf r})]\propto \sum_b \partial_b J_b(\delta a).
	\end{eqnarray}
	However, this is the only closed relation in the constrained problem $\{H_{MF},G, J_b\}$.  All other commutators are not closed. This  means that FVB's procedure cannot be directly applied. 
 {Recently,  Komijani et. al. proposed  a method to solve this problem by including  projectors to impose the constraints in the Hamiltonian~\cite{KKC2023}. This introduces, however, 6-operator interactions.}  {Here, using the BRST invariance as a guiding principle, we  managed to find general gauge fixing conditions that are consistent with both the first and second class constraints.}  {Our approach in principle, is equivalent to the method in~\cite{KKC2023} provided that no further approximations are made. But in practice, various approximations are unavoidable and further research is needed to understand the connections between the approximations used in our work and in~\cite{KKC2023}.}

 {We describe the procedure to determine the gauge fixing conditions in Appendix~\ref{general}. Below we present an intuitive way to find the gauge fixing condition used in our calculations.} 
	 First of all, for any $U(1)$ gauge theory with the temporal and spatial  components of the gauge field, we can start from a well-known gauge fixing condition, the Lorenz gauge,  
	\begin{eqnarray}
		\zeta\partial_\tau\delta\lambda+\sum_b\partial_b \delta a_b=0, \label{LG}
	\end{eqnarray}
	where $\zeta$ is an arbitrary constant introduced for later convenience. 
	With this gauge fixing condition, we have the BRST invariant Lagrangian in the continuous limit 
	\begin{eqnarray}
		L^{(2)}_{BRST}&=&L_{MFC}-\int d^2r\frac{1}{2\xi} (\zeta\partial_\tau\delta\lambda+\sum_a\partial_a \delta a_a)^2\nonumber\\
		&+&\frac1{\xi}\int d^2r\bar u(\zeta\partial_\tau^2+\sum_a\partial^2_a)u, \label{BRSTLG}
	\end{eqnarray}
	where $\xi$ is an arbitrary gauge parameter and the BRST transformations read
	\begin{eqnarray}
		&&\delta_{B}f_\sigma=i\epsilon gu f_\sigma, \delta_{B}h=i\epsilon gu h,\delta_B\phi=2i\epsilon g u,\nonumber\\
		&&\delta_{B} \delta\lambda=\epsilon \partial_\tau u,\delta_B \delta a_b=\epsilon \partial_bu, \delta_{B}u=0,\nonumber\\
		&& \delta_B\bar u=\epsilon (\zeta\partial_\tau\delta\lambda+\sum_b\partial_b\delta a_b);  \label{BRSTT}
	\end{eqnarray}
	and $L_{MFC}$ is the continuum  limit of the mean field Lagrangian~\eqref{LMF}.  {Our theory can be applied to various pairing potentials, but here we focus on the $d$-wave pairing, such that  $L_{MFC}$ takes the following form}  
	\begin{eqnarray}
		L_{MFC}&=&\int d^2r [ \sum_{\sigma}f^\dag_{\sigma}(\partial_\tau-\mu_f-ig \delta \lambda)f_\sigma\nonumber\\
		&+&h^\dag(\partial_\tau-\mu_h-ig \delta \lambda)h]\nonumber\\
		&-&\int d^2r[\frac{1}{2m_f} \sum_{\sigma,a}f^\dag_{\sigma}(-i\partial_a-g \delta a_a)^2f_\sigma\nonumber\\
		&-&\frac{1}{2m_h}\sum_ah^\dag(-i\partial_a-g \delta a_a)^2h]\label{LMFC}\\
		&+&\frac{1}2\int d^2r \sum_a (\Delta_a \partial_a(e^{i\phi/2}f^\dag_\uparrow) \partial_a(e^{i\phi/2}f^\dag_\downarrow)+h.c.).\nonumber
	\end{eqnarray}

However, it is easy to check that the determinant of the free gauge field propagator read out from~\eqref{BRSTLG} is singular, i.e.,  $\det {\cal D}^{-1}(i\nu_n,{\bf q})=0$. This means that the redundant gauge degrees of freedom are not completely fixed. A further gauge fixing condition is required. For the FVB's procedure in Sec.~\ref{subsection_BRST_exact}, $\partial_\tau^2\lambda_i=0$ [Eq.~\eqref{dotpi}] ensures that $\lambda_i$ is not accelerated. In mean field theory,  {inspired by the fact that the spatial gauge fluctuations are included, we generalize this condition into a D'Alembert-like one}, $\partial_\tau^2\delta \lambda+\frac{1}{\xi}\sum_b \partial_b^2\delta\lambda=0$,  {where $\xi$ is a parameter.}  {For a rigorous derivation of this condition, see Appendix~\ref{general} where a general form of the quadratic gauge fixing Lagrangian with BRST symmetry is presented}. By acting $\partial_\tau$ on the Lorenz gauge~\eqref{LG}, we have $\zeta\partial_\tau^2\delta\lambda+\sum_b\partial_\tau\partial_b\delta a_b=0$. The D'Alembert-like condition, combined with this equation, reduces to the following gauge fixing condition  to the  Lagrangian~\eqref{BRSTLG}, 
	
	\begin{eqnarray}
		\frac{\zeta}{\xi}\sum_b \partial_b^2\delta\lambda-\sum_b\partial_\tau\partial_b \delta a_b=0. \label{GF1}
	\end{eqnarray} 
	We find that, up to some total divergence terms, the BRST invariant Lagrangian that  {is} consistent with the constraints $G=0$ and $J_b=0$ as well as the gauge fixing conditions~\eqref{LG} and~\eqref{GF1} is given by   
	
	\begin{eqnarray}
		L_{BRST}&=&L_{MFC}{-}\int d^2r(\frac{\zeta}{2}(\partial_\tau \delta \lambda )^2+\frac{1}{2\xi}(\sum_b \partial_b \delta a_b)^2\nonumber\\
		&+&\frac{1}{2}\sum_b(\partial_\tau \delta a_b)^2+\frac{\zeta}{2\xi}\sum_b (\partial_b \delta \lambda)^2)\nonumber\\
		&+&\frac1{\xi}\int d^2r\bar u(\zeta\partial_\tau^2+\sum_a\partial^2_a)u\nonumber\\
		&\equiv&L_{eff}+\frac1{\xi}\int d^2r\bar u(\zeta\partial_\tau^2+\sum_a\partial^2_a)u.\label{LB2}
	\end{eqnarray}
	The determinant of the free gauge propagator corresponding to~\eqref{LB2} is non-vanishing besides some poles; see Eq.~\eqref{GGFI}.
	
	In this way, we obtain a BRST symmetric theory with second-class constraints. 
	By Noether's theorem, the Euclidean BRST charge is then given by (see Appendix~\ref{general})
	
	\begin{eqnarray}
		Q_B&=&\int \mathrm{d}^2x~ (igG +\frac{\zeta}{\xi}\partial^2\delta\lambda-\sum_b\partial_\tau\partial_b \delta a_b )u\nonumber\\
		&&+[\zeta\partial_\tau\delta\lambda +\sum_b\partial_b \delta a_b]\partial_\tau u.\label{BRSTC}
	\end{eqnarray}
	The physical states are then constrained by~\cite{Weinberg}
	\begin{eqnarray}
		Q_B|{\rm phys}\rangle=0.
	\end{eqnarray}
 Since the local ghost field $u$ and its $\tau$-derivative $\partial_\tau u$  are independent, we recover the constraint $G=0$ and the gauge fixing conditions~\eqref{LG} and~\eqref{GF1}. 
	
	However, the vanishing counterflow condition $J_b=0$ is not included in $Q_B|{\rm phys}\rangle=0$. Notice that to obtain the BRST charge~\eqref{BRSTC} from Noether's theorem, the Euler-Lagrange equations of motion of all fields are already used. Thus, we check the equations of motion  {involved} in the current $J_b(\delta  a)$.  According to the Euler-Lagrange equations for $\delta a_b$
	\begin{eqnarray}
		\frac{\delta L_{eff}}{\delta (\delta a_b)}-\partial_\tau (\frac{\delta L_{eff}}{\delta \partial_\tau\delta a_b})-\sum_c\partial_c (\frac{\delta L}{\delta \partial_c \delta a_b})=0,
	\end{eqnarray}
	we have 
	\begin{eqnarray}
		J_b(\delta  a) -\frac{1}{2}\partial^2_\tau \delta a_b- \frac{1}{\xi}\partial_b(\sum_c\partial_c \delta a_c)=0,\label{Jne0}
	\end{eqnarray} 
	which does not directly give the vanishing spinon-holon counterflow constraint. As we have pointed out, we need to 
	average Eq.~\eqref{Jne0} for the fluctuating gauge field, and the physical gauge invariant current obeys
	\begin{eqnarray}
		&&\langle J_b(\delta  a) -\frac{1}{2}\partial^2_\tau \delta a_b- \frac{1}{\xi}\partial_b(\sum_c\partial_c \delta a_c)\rangle_{\delta a}\nonumber\\&&=\langle J_b(\delta  a)\rangle_{\delta a}=0,
	\end{eqnarray} 
	which is the vanishing spinon-holon counterflow constraint.

	Summarily, in the BRST quantization procedure, the redundant gauge degrees of freedom are fixed while the original physical constraints are consistently  {maintained}.  Based on  {this} well-defined theory, we can  {perform} perturbation calculations.  Again, for the $U(1)$ gauge theory, the ghost part is decoupled from $L_{eff}$, and we  {will only} consider the Lagrangian $L_{eff}$  {for} the  {rest} of this work. 
	
	\subsection {The Higgs Mechanism}\label{SectionIIC}
 
	Before  {proceeding with} perturbation calculations,  {let us} discuss the Higgs mechanism for the pairings.  {By} defining $\psi_\sigma=e^{-i\phi/2}f_\sigma$, the fermionic field $\psi_\sigma$  {remains} gauge invariant. The effective Lagrangian becomes 
	
	\begin{eqnarray}
		L_{eff}&=&\int d^2r\sum_{\sigma}\psi^\dag_{\sigma}[ \partial_\tau-\mu_f-ig{\cal A}_0\nonumber\\
		&-&\frac{1}{2m_f} \sum_{a}(-i\partial_a-g{\cal A}_a)^2]\psi_\sigma\nonumber\\
		&+&\int d^2rh^\dag[\partial_\tau-\mu_h+g \delta \lambda\nonumber\\
		&-&\frac{1}{2m_h}\sum_a(-i\partial_a-g \delta a_a)^2]h\nonumber\\
		&+&\int d^2r\sum_a(\Delta_a \partial_a\psi^\dag_\uparrow\partial_a\psi^\dag_\downarrow)+h.c.)\\
		&-&\int d^2r(\frac{\zeta}{2}(\partial_\tau \delta \lambda )^2+\frac{1}{2\xi}(\sum_b \partial_b \delta a_b)^2\nonumber\\
		&+&\frac{1}{2}\sum_b(\partial_\tau \delta a_b)^2+\frac{\zeta}{2\xi}\sum_b (\partial_b \delta \lambda)^2).\nonumber
	\end{eqnarray} 
 We see that  {the} $\phi$ field is absorbed into $\psi$, which is gauge invariant, while $g{\cal A}_0=g\delta \lambda-\dot \phi/2$ and $g{\cal A}_a=g\delta a_a-\partial_a \phi/2$ 
	are also  gauge invariant.  This is the Higgs mechanism  {and it is known that no further gauge fixing condition is required.}  {To see this explicitly, we check the equation of motion of $\phi$.} Varying $\phi$, we obtain
	\begin{eqnarray}
		\partial_\tau J_{\psi \tau}-\partial_bJ_{\psi b}=0.
	\end{eqnarray}
	where $J_{\psi \tau}=\sum_\sigma \psi^\dag_\sigma \psi_\sigma=n_f$ and $J_{\psi b}=-\frac{1}{m_f}\sum_\sigma \psi_\sigma^\dag(i\partial_b+g {\cal A}_b)\psi_\sigma=J_{fa}$.  This is exactly the spinon current conservation and does not result in a new constraint.  In the other words, $\phi$ is not a gauge field, and there  {are} no redundant degrees of freedom  to be fixed. The effect of $\phi$  {on} the pairing physics will be studied in other works.

	\section{The perturbation theory} \label{sectionIII}
	
	We study the strange metal phase where the holons are not condensed and the spinons are not paired.  Since $\Delta_a=0$, we do not need to distinguish $f_\sigma$ from $\psi_\sigma$, i.e., we take $\phi=0$. 
	According to the Faddeev-Popov path integral quantization, the gauge invariant partition function with no redundant gauge degrees of freedom is given by
	\begin{eqnarray}
		Z_{eff}&\propto& \int \prod_{{\bf r},\tau}d\Phi d\bar u du e^{-\int_0^\beta d\tau L_{BRST}}\nonumber\\
		&\propto& \int \prod_{{\bf r},\tau}d\Phi e^{-\int_0^\beta d\tau L_{eff}}.
	\end{eqnarray}

	\subsection {Non-interacting Green's functions and interaction vertices}

 
	We  {can} now draw Feynman's diagrams according to $L_{eff}$. 
	Taking $\mu=\tau,1,2$ and $k_\mu=(\nu_n, k_{1}, k_2)$, the inverse of the free one-particle Green's function of the gauge field is given by
	\begin{eqnarray}
		&&	{\cal D}^{(0)-1}_{\mu\nu}({\bf k},i\nu_n)=\nonumber\\
		&&	-\left[
		\begin{array}{cccc}
			\zeta\nu^2_n+\frac{\zeta}{\xi}\mathbf{k}^2 &0 & 0\\
			0 & \nu^2_n+\frac{1}{\xi}k^2_x & \frac{1}{\xi}k_x k_y\\
			0 & \frac{1}{\xi} k_x k_y & \nu^2_n+\frac{1}{\xi}k^2_y
		\end{array}
		\right].
	\end{eqnarray} 
	The determinant of this matrix is
	\begin{eqnarray}
		\det ({\cal D}^{(0)-1}_{\mu\nu})=\zeta\nu^2_n(\nu_n^2+\frac{1}{\xi}{\bf k}^2)^2. \label{GGFI}
	\end{eqnarray}
	As expected, the Green's function is regular  except for some singular poles. Besides $\nu^2_n=0$, we have two poles at $\nu^2_n=-{\bf k}^2/\xi$. 
	
	The one-particle Green's function of the gauge field is then given by 
	\begin{eqnarray}
		&&	{\cal D}^{(0)\mu\nu}({\bf k},i\nu_n)=-\frac{1}{\nu^2_n(\nu^2_n+{\bf k}^2/\xi)}\nonumber\\
		&&	\left[
		\begin{array}{cccc}
			\frac{1}{\zeta}\nu^2_n &0 & 0\\
			0 & \nu^2_n+({\bf k}^2-k^2_x)/\xi & -k_x k_y/\xi\\
			0 & -k_x k_y/\xi & \nu^2_n+({\bf k}^2-k^2_y)/\xi 
		\end{array}
		\right].\label{GGF}
	\end{eqnarray} 
	
\begin{figure}
		\includegraphics[width=0.4\textwidth]{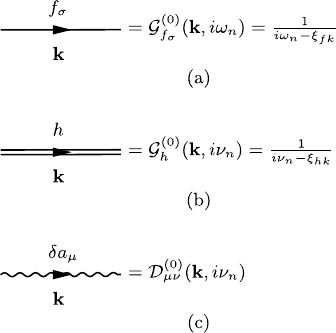}
		\caption{
			{ The Feynman diagrams for the free one-particle Green's functions of (a) $f_\sigma$, (b) $h$, (c) $\delta  a_\mu=(\delta\lambda,\delta a_a)$. $\xi_{f(h)k}=\frac{k^2}{2m_{f(h)}}-\mu_{f(h)}$. }
			\label{fig1}	}
	\end{figure}

	The one-particle Green's functions of the other fields can be easily read out from $L_{eff}$.   The free one-particle Green's functions for $\psi_\sigma, h,\delta\lambda$ and $\delta \vec a$ are shown in turns in Fig. \ref{fig1}.

	When $h$  {condenses}, the holon Green's function becomes $G_h=\rho_{h0}+G'_h$. In the  {fermion-paired} phases, there exist anomalous Green's functions. In this work, we do not intend to deal with the paired states and thus put these Green's functions in Appendix~\ref{AAn}.

	
 {The interaction vertices can be directly read out from $L_{eff}$; see Fig.~\ref{fig2} and Fig.~\ref{fig3}. The Feynman's rules and Dyson's equations are given in Appendix~\ref{feynmanRule}.}
	
	\begin{figure}	
		\includegraphics[width=0.4\textwidth]{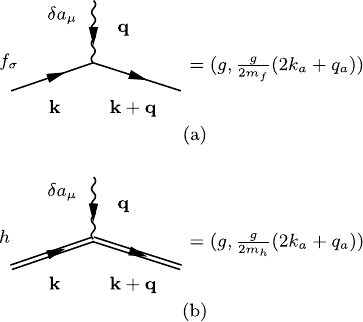}
		\caption{
			{ The Feynman diagrams for the 3-point vertex. }
			\label{fig2}	}
	\end{figure}

	\begin{figure}
		\includegraphics[width=0.3\textwidth]{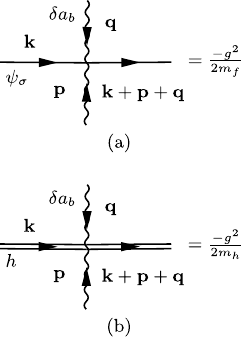}
		\caption{{ The Feynman diagrams for the 4-point vertex. }\label{fig3}}
	\end{figure}

	\section{One-electron Green's function and non-fermi liquid}\label{sectionIV}
	
	In this section, we  {will} calculate the one-electron Green's function and the momentum distribution of the electron. 

	\subsection{One-electron Green's function}

\begin{figure}
		\includegraphics[width=0.35\textwidth]{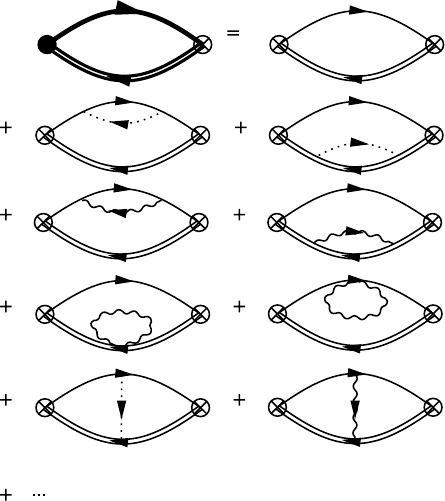}
		\caption{{The electron Green's function is constructed from the full Green's functions of holon and spinon, and a full vertex correction, which are  {represented} by thick lines and black circles.  {From the second to the fourth lines represent}   the self-energy corrections of  {holons} and spinons  {up} to $g^2$ order. The fourth line  {represents} the vertex correction  {up} to  $g^2$ order. The $\cdots$  {represents} higher order terms. The Green's function of $\delta\lambda$ is  {indicated} by the dotted line, while the wavy line  {indicates} that of $\delta a_b$. }
			\label{fig4}	}
	\end{figure}

	The one-electron Green's function in the slave boson theory is given by 
	\begin{eqnarray}
		G_{e\sigma}({\bf r},\tau)=-\langle T_\tau(h^\dag({\bf r},\tau)f_\sigma ({\bf r},\tau)f^\dag_\sigma(0,0)h(0,0))\rangle. 
	\end{eqnarray}
	In the path integral calculation, we introduce a fermionic source term to obtain the electron Green's function through the function derivative
	\begin{eqnarray}
		G_{e\sigma}({\bf r},\tau)=-\frac{\delta^2 W_\eta}{\delta\bar \eta({\bf r},\tau)\delta\eta({\bf r},\tau)}\bigg|_{\bar\eta=\eta=0},\label{OEG}
	\end{eqnarray}
	where the free energy $W_\eta$ is defined by  
	\begin{eqnarray}
		&&e^{-W_\eta}\equiv Z_{eff,\eta}=\int \prod_{{\bf r},\tau}d\Phi({\bf r},\tau) \exp\{-\int d\tau L_{eff}\nonumber\\
		&&-\int d\tau d^2r( \bar\eta({\bf r},\tau)h^\dag({\bf r},\tau)f_\sigma({\bf r},\tau)+\eta({\bf r},\tau)h({\bf r},\tau)f^\dag_\sigma({\bf r},\tau))\}\nonumber.
	\end{eqnarray}
	The second line is called the source vertex, which can be diagrammatically represented as
 	\begin{eqnarray}
		\otimes=-\int d\tau d^2r \eta ({\bf r},\tau).
	\end{eqnarray}
	The electron Green's function can then be calculated by the Feynman diagram; see Fig.~\ref{fig4}.
	
	We remark that the one-electron Green's function is gauge invariant. According to the perturbation theory, this gauge invariance is kept  loop by loop.  However, under the RPA, the results may be gauge-dependent because we have neglected the fluctuations from the source vertex and gauge field self-energy. Due to the  {complexity} of the full Green's function, we here perform the RPA calculation and leave the gauge invariant calculation to further work.

	\subsection{Momentum distribution}
	
	According to the electron Green's function~\eqref{OEG}, one can calculate the momentum distribution of the electrons. Without considering  gauge fluctuations,  the Green's function~\eqref{OEG} has been calculated in~\cite{AZ,Rodriguez}. 
	
	The momentum distribution of the electrons is defined by
	\begin{eqnarray}
		n_{\bf k}=2\langle c^\dagger_{\sigma {\bf k}}c_{\sigma {\bf k}}\rangle
		=2\sum_{{\bf q,q'}}\langle f^\dag_{\sigma {\bf k+q}}   f_{\sigma {\bf k+q'}}h_{\bf q}h^\dag_{\bf q'} \rangle.
	\end{eqnarray}
	The momentum distribution may also be calculated according to the spectral function, i.e., 
	\begin{eqnarray}
		n_ {\bf k}&=&\int \frac{d\omega}{2\pi} A_e({\bf k},\omega)n_F(\omega), 
	\end{eqnarray}
	where $n_F(\omega)$ is the Fermi distribution.  The electron spectral function $A_{e} ({\bf k},\omega)=-2 {\rm Im} G^r_{e\sigma}({\bf k},\omega)$.  
	
	We calculate $G^r_{e\sigma}({\bf k},\omega)$ perturbatively.  To the zeroth order of $g$, there is no correction to the source vertex, and the single electron Green's function is approximated by the right diagram in the first line of Fig.~\ref{fig4}, which in the  {Matsubara} frequency space is
	\begin{eqnarray}
		{\cal G}^{(0)}_{e\sigma}(\mathbf{k},i\omega_n)&=&\frac{1}{N\beta}\sum_{m}\int\frac{\mathrm{d^2}q}{(2\pi)^2} {\cal G}^{(0)}_{f\sigma}(\mathbf{k}+\mathbf{q},i\omega_n+i\nu_m)\nonumber\\
		&\times&{\cal G}^{(0)}_h(\mathbf{q},i\nu_m),\label{oeg}
	\end{eqnarray}
	where $N$ is the number of lattice sites. This expression has been obtained in~\cite{AZ}.  The retarded thermodynamic  {Green's} function $G^r_{e\sigma}({\bf k},\omega)$ is given by $i\omega_n\to \omega+i0^+$ in the Matsubara function. Thus, the  {zeroth-order} contribution to the momentum distribution is given by 
	\begin{eqnarray}
		n_{\bf k}^{(0)}&=&\frac{1}{(2\pi)^3}\int d^2q \int d\omega n_F(\omega)A^{(0)}_f({\bf k+q},\omega)\nonumber\\&&(1+\int d\nu n_B(\nu)B^{(0)}_h({\bf q},\nu)),
	\end{eqnarray}
	where $A^{(0)}_f=-2{\rm Im}G^{r(0)}_{f\sigma }$ is the free spinon spectral function and $B^{(0)}_h=-{\rm Im}G^{r(0)}_h$ is the free holon spectral function with $G^{r(0)}_h$ being the retarded free Green's function of the holon, 	and $n_{F/B}$ is the fermion/boson distribution function at temperature $T$.  {In} the strange metal phase, there is no holon condensation. Thus, at zero temperature, 
	\begin{eqnarray}
		n_{{\bf k},T=0}^{(0)}=\frac{1}{(2\pi)^3}\int d^2q \int d\omega \Theta(-\omega)A^{(0)}_f({\bf k+q},\omega), 
	\end{eqnarray}
	where $\Theta(x)$ is the step function. It is clear that the  momentum distribution $n_{{\bf k},T=0}^{(0)}$ is a constant independent of ${\bf k}$. If we do not consider the  {higher-order} contributions, the momentum distribution obeys the sum rule 
	\begin{eqnarray}
		1-x=\frac{1}{L^2}\int \frac{d^2k}{(2\pi)^2} n_{\bf k} \label{sr},
	\end{eqnarray}
	which gives $n_{{\bf k},T=0}^{(0)}\sim 1-x$ at zero temperature for $L$ being the lattice size.  Furthermore, it is easy to see that  {by} replacing the free spinon and holon Green's functions  {with} the full ones and neglecting  the vertex correction, the zero temperature momentum distribution is still constant. 
	This is an extreme non-Fermi liquid behavior because although the momentum distribution of the spinon obeys the standard Fermi distribution, the integrations over $\omega$ and $\mathbf{q}$  {destroy} the discontinuity at the Fermi momentum of  the electron momentum distribution. However, if we do not consider the source vertex correction, this constant momentum distribution is obviously not physical.   
	Thus, to obtain a physical result, the spinon-holon source vertex correction with the effect of the dynamics of the gauge field has to be considered. 
	\begin{figure}
		\vspace{3mm}
		\includegraphics[width=0.4\textwidth]{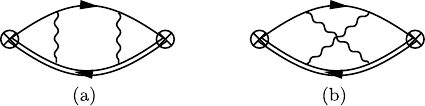}
		\caption{
			{ The Feynman diagrams of the $g^4$ order with ${k_ak_b}$-dependent contributions. }
			\label{fig5}	}
	\end{figure}

	The $g^2$ order correction to the source vertex comes from the fourth line of Fig.~\ref{fig4}. But their contribution to the momentum distribution is still a constant, i.e., independent of the momentum. See Appendix.~\ref{momentum_dis} for details. Let us check the $g^4$ order contribution. Besides the linearly $\bf k$-dependent diagrams, which eventually become zero due to reflection symmetry, the diagrams in Fig.~\ref{fig5} provide the nonzero quadratic ${ k_ak_b}$-dependent contributions to the electron momentum distribution. And we have (see Appendix.~\ref{momentum_dis} for details) 
	\begin{eqnarray}
		n^{(4)}_{e{\bf k},T=0}=2(n_{e\sigma\bf k}^{4A}+n_{e\sigma\bf k}^{4B})=
    -C^{(4)}k^2.
	\end{eqnarray}
	Other contributions to the $k^2$ terms come from the correction of ${\cal G}_{00}$ to $n^{(4)}_{e{\bf k},T=0}$, which are of the order $O(g^6)$ and higher. Taking these contributions into account,  $C^{(4)}$ is corrected to $C^{(4)}$. Similarly, the $2n$-loop diagrams with the spinon-holon vertex  {and} $2n-1$ lines of the spatial gauge field Green's function do not contribute to the $g^{4n-2}$ order,  while such $2n+1$  {loop} diagrams contribute to $n_{e\bf k}$ with
	\begin{eqnarray}
		n^{(2n+2)}_{e{\bf k}, T=0}=(-1)^nC^{(2n+2)} k^{2n},
	\end{eqnarray}  
	and $C^{(2n+2)}$ is corrected to $\tilde C^{(2n+2)}$. The momentum distribution at $T=0$ is then given by
	\begin{eqnarray}
		n_{e{\bf k},T=0}=\sum_{n=0} (-1)^n\tilde C^{(2n+2)} k^{2n}.
	\end{eqnarray}
 
	This is a continuous function of $k^2$ and there is no  {jump} at the Fermi momentum $k_F$. 
 Note that $n_{e{\bf k=0},T=0}$ does not  {approach} unity when ${\bf k}\to 0$. The discontinuity of the spinon momentum distribution  {is} integrated to be smooth, and the $n_{e{\bf k},T=0}$ near $k\sim k_F$ is not an interaction-dependent  power law like $|k-k_F|^{\alpha(J)}$ for an $\alpha$ depending on $J$ as in the Luttinger liquid. Instead, there is no particularity about the expansion at $k_F$. In general, the expansion around any given $k_0$ is linear in $|k-k_0|$ based on loop expansion in Feynman diagrams. 
 {We note that Luttinger's theorem~\cite{Lutt}, although has a topological origin and is valid for certain non-Fermi liquids~\cite{Oshikawa2000}, can be violated in strongly correlated systems~\cite{Goldman_2023}
 . In our theory,  Luttinger's theorem does hold for spinons.  {However}, since the one-electron Green's function is actually the spinon-holon two-particle Green's function and the holons do not condense, Luttinger's theorem breaks down for electrons. The violation of  Luttinger's theorem, however, should be viewed with caution because we have not  {proven} that our perturbative expansion converges. The possibility of the breakdown of  Luttinger's theorem will be explored in the future.}
 We  do not give the numerical result for $n_{e{\bf k},T=0}$  {here} because the  {zero-temperature} momentum distribution of the strange metal  is not  {experimentally} observable. Instead, we calculate the  {finite-temperature} electron spectral function which is experimentally more relevant.

	\subsection{The electron spectral function} \label{specf}
 
	We showed that without  the source vertex correction, the momentum distribution at zero temperature is constant. However, the spectral function without the source vertex correction depends on ${\bf k}$ and $\omega$ at  {a} finite temperature. In this section, we focus on the finite temperature case and neglect the source vertex correction and use the RPA correction to Eq.~\eqref{oeg} from the spinon and holon self-energies to calculate $A_e({\bf k},\omega)$. 
	According to the Dyson equations (see Appendix~\ref{feynmanRule}), the one-electron  {Matsubara}'s function in this approximation can be written as 
	\begin{eqnarray}
		&&{\cal G}_{e\sigma}({\bf k},i\omega_n)\nonumber\\
		&=&\frac{1}{\beta(2\pi)^2N}\sum_{m}\int d^2q\frac{1}{i\nu_m-\xi_{h,\bf q}-\Sigma_h({\bf q},i\nu_m)}\nonumber\\
		&\times&\frac{1}{i\omega_n+i\nu_m-\xi_{f,\bf k+q}-\Sigma_f({\bf k+q},i\omega_n+i\nu_m)},
		\label{MEG}
	\end{eqnarray}
	which is also related to the spectral function by 
	\begin{eqnarray}
		{\cal G}_{e\sigma}(\mathbf{k},i\omega_n)=\int\frac{\mathrm{d}z}{2\pi}\frac{A_e(\mathbf{k},z)}{i\omega_n-z}.
	\end{eqnarray}
	
     If one ignores the spinon and holon self-energies in the electron's Green's function~\eqref{MEG}, then the corresponding spectral function was discussed by Lee and Nagaosa~\cite{LN}. They pointed out that the holon part leads to a peak centered around $\mu_f-|\mu_h|$, and the spinon part contributes to a continuum dip-bump with a threshold. They further used a phenomenological gauge propagator to estimate the effect of gauge fluctuations. Here, we use our controlled gauge theory to perturbatively calculate the electron spectral function in the presence of gauge fluctuations. 
	
	With the help of the spinon and holon spectral functions $A_{f,h}(\mathbf{k},z)$, one can write Eq.~\eqref{MEG} as
	\begin{eqnarray}
		&&{\cal G}_{e\sigma}(\mathbf{k},i\omega_n)=-\sum_m\int\frac{d^2q}{(2\pi)^2}\int\frac{dz_1}{\pi}\frac{A_{f\sigma}(\mathbf{k+q},z_1)}{i\omega_n+i\nu_m-z_1}
		\nonumber\\ &&\times\int\frac{dz_2}{\pi}\frac{A_h(\mathbf{q},z_2)}{i\nu_m-z_2}=\int\frac{d^2qdz_1dz_2}{4\pi^4}
		\frac{n_F(z_1)+n_B(z_2)}{i\omega_n-z_1+z_2}\nonumber\\ &&\times A_{f\sigma}(\mathbf{k+q},z_1)A_h(\mathbf{q},z_2).
	\end{eqnarray}
The electron spectral function  {can} then  be expressed  by using spinon and holon spectral functions as
	\begin{eqnarray}
		A_{e\sigma}(\omega,\mathbf{k})&=&
		\int\frac{d^2q dz}{(2\pi)^2\pi}
		[n_F(\omega+z)+n_B(z)]\nonumber\\
		&\times&A_{f\sigma}(\mathbf{k+q},\omega+z)A_h(\mathbf{q},z).
	\end{eqnarray}

	\begin{figure}[t]
		\includegraphics[width=0.5\textwidth]{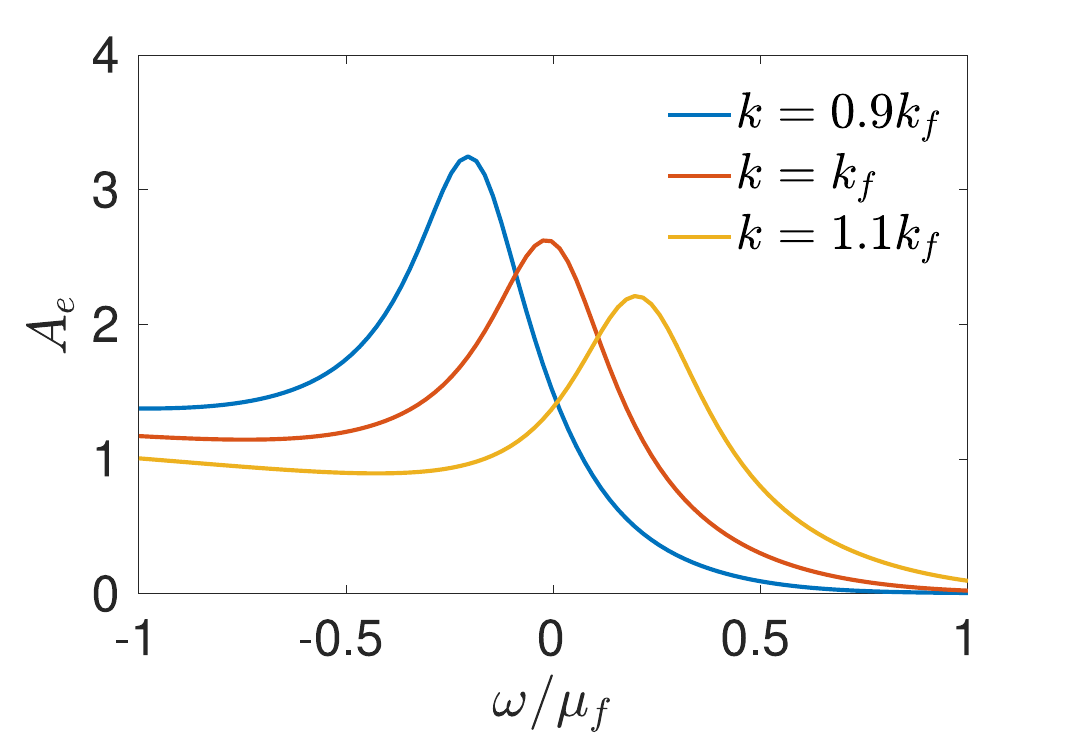}
		\caption{(color online) {The electron spectral function without gauge fluctuations  at $T=200K>T^*$. }}
		\label{fig6}	
	\end{figure}

	We first calculate the electron spectral function with the free spinon and holon by taking $\Sigma _f=0$ and $\Sigma_h=0$ in Eq.~\eqref{MEG}. 
	Without gauge fluctuations, the spinon and holon spectral functions are
	\begin{eqnarray}
		A^0_{f}(\omega,\mathbf{k})=\pi\delta(\omega-\xi_{f,\mathbf{k}}),\\
		A^0_{h}(\omega,\mathbf{k})=\pi\delta(\omega-\xi_{h,\mathbf{k}}),
	\end{eqnarray}
	and then the electron spectral function is given by
	\begin{eqnarray}
		A^0_{e\sigma}(\omega,\mathbf{k})&=&\int\frac{d^2q}{4\pi}
		[n_F(\xi_{f,\mathbf{k}+\mathbf{q}})+n_B(\xi_{h,\mathbf{k}})]\nonumber\\
		&\times&\delta(\omega+\xi_{h,\mathbf{k}}-\xi_{f,\mathbf{k}+\mathbf{q}}). \label{FSH}
	\end{eqnarray}
	We consider a square lattice model with dispersion $E_\mathbf{k}=-t(\cos k_x a+\cos k_y a)-\mu$ with $t\sim0.1 $eV, $\mu\sim -0.05$eV, and $x=0.2$, which are typical for the cuprates in the strange metal phase~\cite{cup}. The lattice constant $a$ is set to be unity. In the continuum limit, the above parameters correspond to $k^2/(2m_f)-\mu_f$ with $a^2t\sim 1/m_f$, and $\mu_f=0.15eV$. The Fermi momentum is $k_F\sim \sqrt3/a$. Near the Fermi surface ($k\approx k_F$), the electron spectral functions with the free spinon and holon are shown in Fig.~\ref{fig6}, in which $T=200K$ is greater than the temperature $T^*$, above which the system is in the strange metal phase. Although this electron spectral function without gauge fluctuations is not physical at zero temperature, it shows some features  {that} also appear in the spectral function with gauge fluctuations. In particular,  we can see the peak of the quasiparticle spectral weight near the Fermi momentum and a dip-bump structure of the spectral function below the Fermi momentum. 
 These features of the electron spectral function  {seem to} fit with the experimental observation at $T\sim 200 K$ as noticed by Anderson and Zou~\cite{AZ}.

We now consider the effect of the gauge fluctuations from the RPA. 
The coupling constant $g$ is taken to be $0.1$ for the perturbation theory, and the gauge parameter is taken to be $\xi=1$. 
In the PRA calculation, we need the one-loop spinon self-energy
	\begin{eqnarray}
		\Sigma^{(0)}_{f}(i\omega_m,\mathbf{q})&=&\sum_{n}\int d^2k~\gamma_{\mu}{\cal D}^{(0)}_{\mu\nu}(i\nu_n,\mathbf{k})\nonumber\\
		&\times&\gamma_\nu {\cal G}^{(0)}_f(i\nu_n+i\omega_m,\mathbf{k+q}),
	\end{eqnarray}
where $\gamma_\mu=-g(i,\frac{\mathbf{k}+\mathbf{q}/2}{m_f})$ are the interaction vertices read from Figs.~\ref{fig2} and~\ref{fig3}. The expression for the holon self-energy is similar, and  detailed expressions are presented in Appendix~\ref{SHSelfEnergy}.  {The momentum integral diverges in $\Sigma_{f,h}^{(0)}$.  The ultraviolet divergence may be removed by simply taking a cut-off because the lattice spacing is finite and we take $k_{UV}=10k_f$ in our numerical calculations.
 {In principle,} the infrared divergence  should be cancelled by other gauge fluctuations, such as the source vertex correction and the gauge field self-energy. Under the RPA, we simply take a long wavelength cut-off  {of} $k_{IR}=10^{-5}k_f$. We checked that the result is not sensitive to the cutoff.
}

	\begin{figure}[h]
		\includegraphics[width=0.5\textwidth]{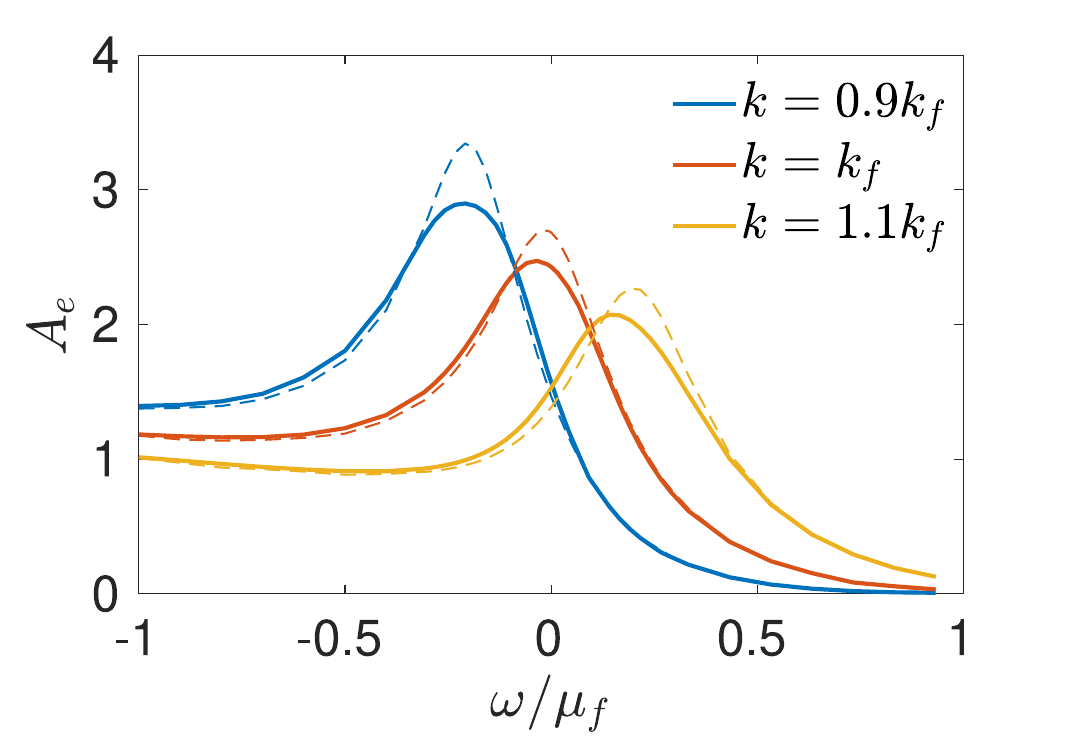}
		\caption{The electron spectral function for different momentum and frequency. Solid line: holon self-energy is neglected. Dashed line: both spinon and holon self-energies are neglected.}\label{fig7}
	\end{figure}

	\begin{figure}[h]
		\includegraphics[width=0.5\textwidth]{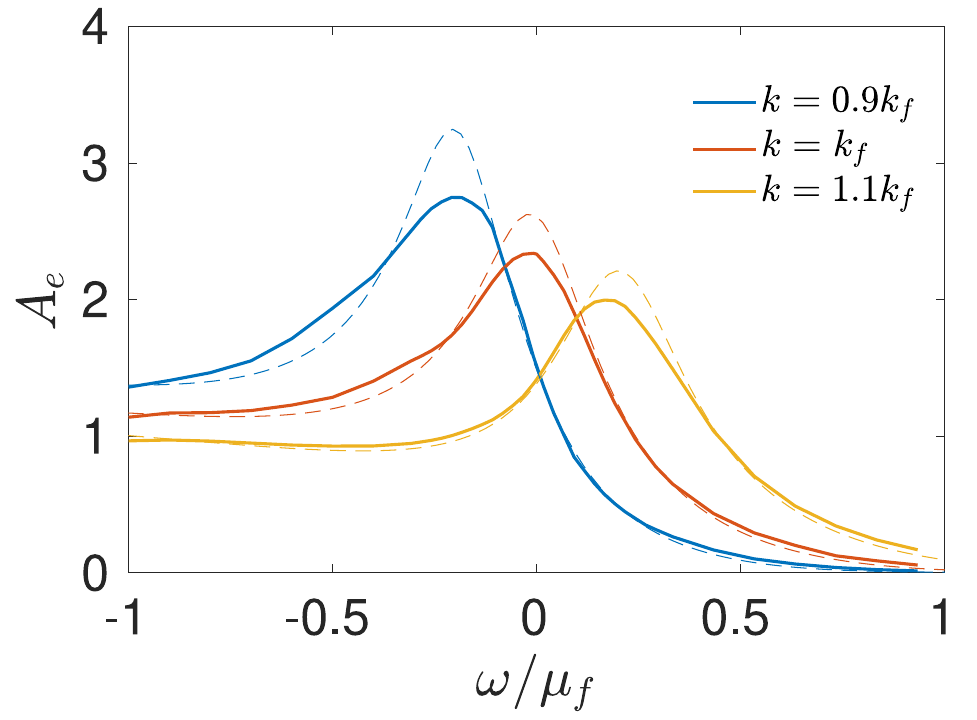}
		\caption{The electron spectral function for different momentum and frequency. Solid line: with gauge fluctuations. Dashed line: without gauge fluctuations. }\label{figAe}
	\end{figure}
	 {Fig.~\ref{fig7} shows the spectral function with only  {spinon} self-energy included, and} in Fig.~\ref{figAe}, we put the gauge fluctuations from both spinon and holon self-energies into the electron spectral function.  {The results are similar.} One can see that gauge fluctuations suppress the peaks of the spectral weight. Our result is consistent with  {the} ARPES measurement in the cuprates. The electron spectral function at $k_F$ is proportional to $\pm \omega$ near the Fermi surface, and the  {slope} vanishes at $k=k_F$, which are generic  {properties} seen in the ARPES measurements for the  {cuprates}. For $k<k_F$, the peak of  {the} spectral function moves inward and becomes higher, and for $k>k_F$, it moves outward and becomes lower, and its $1/\omega^2$ decay away from the Fermi surface is also a characteristic feature of the spectral function. Furthermore, the dip-bump of the spectral weight was also observed in experiments. For  {a} review of the ARPES experiments for the cuprates, see, for example, references~\cite{ARPES1, ARPES2,ARPES}. 
	
	In principle, the BRST symmetry guarantees the gauge invariance of the theory and the physical correlation functions. 
 But in practice, correlation functions may become  {gauge-dependent} because approximations cannot be avoided in the perturbation computation. To obtain the  {gauge-invariant} form of correlation functions, a Ward identity, which is the quantum version of the conservation law, is required. As long as the path integral measure does not change under  {the} BRST transformation,  there is no quantum anomaly. According to the general theory of gauge  {fields}, if we consider all diagrams in the same order of  perturbation, the Ward identity is automatically satisfied. However, we have used  {the} RPA method when calculating the spectral function,  {so} the results will be  {gauge-dependent} in general. The proof of the Ward identity and renormalizability is now beyond the goal of this work, and we shall present them in future works. 
 To check the effect of the gauge dependence, we take another gauge choice, and the electron spectral function behaves similarly to that shown in Fig.~\ref{figAe}.

 {Before closing this section, two  remarks are in order: 
 (i) The conventional wisdom is that the ARPES data is proportional to the electron spectral function, which is dependent on the coupling constant $g$, and we take $g=0.1$ in our calculations. The interpretation of the ARPES data in terms of the spectral function $A_e({\bf k},\omega)$ is based on the sudden approximation~\cite{XiangWu_2022}, which introduces the gauge coupling constant in the photoemission current. In this approximation, the coupling constant can be determined phenomenologically. Precisely, the photoemission current can be perturbatively  calculated by evaluating the correlation function of three current operators, and the result does not depend on the coupling constant $g$~\cite{XiangWu_2022}.} We will give an example to demonstrate this kind of independence in Sec.~\ref{Coupling}. 
 (ii) We also notice that when the rotational symmetry reduces to the $C_4$ symmetry for the square lattice, the spectral function becomes angle-dependent, as found in the ARPES measurement. We will leave these further calculations and the detailed comparisons to the experiments to  further works.

\section{ responses to the external electric and magnetic fields}\label{sectionV}

	\subsection{Ioffe-Larkin rule and the coupling constants}\label{Coupling}

 
	We now study the linear responses to the external electromagnetic field in the strange metal phase.  
	We first review the Ioffe-Larkin composite rule \cite{IL}. In the original $t$-$J$ Hamiltonian, the external electromagnetic field couples to  {the} $t$-term
	\begin{eqnarray}
		tc^\dag_ic_j+h.c.\to e^{iA_{ij}}c^\dag_ic_j+h.c.\to e^{iA_{ij}}f^\dag_if_jh_ih^\dag_j+h.c..~~
	\end{eqnarray}
	This means that in the electromagnetic $U(1)$ gauge transformation $A_{ij}\to A_{ij}+\vartheta_i-\vartheta_j$, the spinon's and holon's gauge transformations transform the $t$-term as  $(f^\dag_if_j)(h_ih^\dag_j)\to(e^{i\alpha\vartheta_{ij}}f^\dag_if_j)(e^{-i\beta\vartheta_{ij}}h_ih^\dag_j)$ for $\alpha-\beta=1$. 
	We chose $\alpha=1$ and $\beta=0$, and physically this means that the electromagnetic field couples to the spinon only~\cite{WNL}. Thus, if $E_a$ is the external electric field and $E_{in}^a$ is the `electric' field for $\delta a_a$, the  currents are
	\begin{eqnarray}
	J_f^a=\sigma_f(E^a+\frac{g}{e}E_{in}^a), J_h^a=\sigma_h \frac{g}{e}E_{in}^a.
	\end{eqnarray}
     {In the above equations the gauge coupling $g$ and electric charge $e$ are written explicitly.}
	Using the constraint $J_f^a+J_h^a=0$, we have 
	\begin{eqnarray}
		E_{in}^a=-\frac{\sigma_f}{\sigma_f+\sigma_h}E^a.
	\end{eqnarray}
	The electric current coincides with the spinon current
	\begin{eqnarray} 
		J^a=J^a_f=-J^a_h=\frac{\sigma_h\sigma_f}{\sigma_f+\sigma_h}E^a=\sigma E^a, \label{JA}
	\end{eqnarray}
	which gives the Ioffe-Larkin composite rule
	\begin{eqnarray}
\sigma^{-1}=\sigma^{-1}_h+\sigma^{-1}_f, R=R_f+R_h.
	\end{eqnarray}
	This means that the spinon and holon form a sequential circuit, not a parallel one~\cite{IL}.  {Note that the gauge coupling constant $g$ does not appear in the Ioffe-Larkin rule, in other words,  the Ioffe-Larkin rule is satisfied no matter what the value of $g$ is.} 
 



 
	According to the linear response Kubo formula and the constraints $J_{f,\mu}=-J_{h,\mu}$ (where $J_{h\tau}=-1-h^\dag h=-hh^\dag$), we have
	\begin{eqnarray}
		J_\mu({\bf q},\omega)=\Pi_{e,\mu\nu}({\bf q},\omega)A^\nu=J_{f,\mu}({\bf q},\omega)=-J_{h\mu}({\bf q},\omega).~~
	\end{eqnarray}
	This gives that 
	\begin{eqnarray}
		\Pi^{-1}_{e,\mu\nu}({\bf q},\omega)= \Pi^{-1}_{f,\mu\nu}({\bf q},\omega)+\Pi^{-1}_{h,\mu\nu}({\bf q},\omega).
	\end{eqnarray}
	We now calculate $\Pi^{f,h}_{ab}({\bf q},\omega)$ using perturbation theory. For expressions for  {the} polarization function, see Appendix~\ref{Appendix:Polarization}.

	\subsection{ Linear-dependence on $T$ of resistivity}


We first qualitatively estimate these electromagnetic responses. We focus  {on} the 
resistivity in the high temperature limit. The spinon contribution to the conductivity is $\sigma_f\sim \Pi_{faa}/\omega$ with $\Pi_{faa}$ being the diagonal polarization, and we find
	\begin{eqnarray}
		\sigma_f^{-1}\sim \frac{\omega}{B_{f}+A_fT},
	\end{eqnarray}
 where the coefficients $A_{f}$ and $B_{f}$ are temperature-independent.  {Similarly}, the holon contribution to the conductivity is 
 \begin{eqnarray}
     		\sigma_h^{-1}\propto 
       \frac{\omega}{x}T.
 \end{eqnarray}
	This is the same result obtained in~\cite{NLPRL,LN}. 
	As recognized by  Nagaosa and Lee~\cite{NLPRL,LN}, the spinon conductivity dominates, i.e., $\sigma_h^{-1}\gg \sigma_f^{-1}$.  {We} then have 
	\begin{eqnarray}
\sigma^{-1}=\sigma^{-1}_f+\sigma^{-1}_h\approx \sigma^{-1}_h,
	\end{eqnarray}
	which is linearly dependent on $T$. Note that our result is obtained  {at} high temperature. Experimentally, the $T$ linear  {behavior} persists down to the superconducting transition temperature and has  {Planckian} slope~\cite{Planckian_diss}, which  {goes} beyond the scope of this article and will be investigated in future work.

	\subsection{Hall resistivity }\label{Hall_LowT}


	We have neglected the terms  {that} are proportional to $q^2$ of the polarization in the last subsection. The coefficients before $q^2$ are known as the Landau diamagnetic susceptibilities for  {spinon} and holon, which can be calculated through the momentum expansion of the polarization functions in the zero frequency limit,
    \begin{eqnarray}\label{chif}
		\chi_f&=&\frac{g^2}{8m_f^2\pi^2N}\int d^2p \frac{n_F(\xi_{f,\bf p+q})-n_F(\xi_{f,\bf p})}{\xi_{f,\bf p}-\xi_{f,\bf p+q}},
	\end{eqnarray}
 and 
 	\begin{eqnarray}
		\chi_h&=&\frac{g^2}{16m_h^2\pi^2 N}\int d^2p
		\frac{n_B(\xi_{h,\bf p+q})-n_B(\xi_{h,\bf p})}{\xi_{h,\bf p}-\xi_{h ,\bf p+q}}.\label{chih}
  \end{eqnarray}
 The Hall resistivity is then given by~\cite{NLPRL,LN}, 
	\begin{eqnarray}\label{HallR}
		R_H=\frac{R_{f,H}\chi_h+R_{h,H}\chi_f}{\chi_h+\chi_f},
	\end{eqnarray}
 where $R_{f,H}$ and $R_{h,H}$ are the Hall resistivity of the  spinon and holon, respectively, and $\chi_f$ and $\chi_h$ are the Landau diamagnetic susceptibilities for the spinon and holon.  
 In the low temperature limit, we found that $\chi_f\sim (1-x)/m_f$ and $\chi_h\sim 2\pi x /m_hT$, which are  {the} same as the results obtained in~\cite{NLPRL,LN}. 
 Taking $R_{h,H}\approx 1/x$, $R_{f,H}\approx -1/(1-x)$, 
 the Hall resistivity of the free spinon and holon \cite{NLPRL,LN}, the Hall resistivity (\ref{HallR}) in the low temperature limit is approximated by~\cite{NLPRL,LN}
 \begin{eqnarray}
     R_H\approx -\frac{1}{1-x}+\frac{1}{x (1-x+\frac{2\pi  m_f x}{m_h T})},\label{Eq:HallR_LT}
 \end{eqnarray}
 which increases as temperature  {rises}. However, the temperature dependence is opposite to the experimental measurements~\cite{Fu,Hall1,Hall2}.  
 	
	In a previous theory, Chien $et$ $al$. introduced an additional scattering time to explain the Hall coefficient anomaly in cuprates~\cite{chien}. However, the origin of such an additional scattering time was not found in the gauge theory. We now see if there is  {a} Hall  coefficient anomaly in our theory.  We do not plan to  {examine} if the  {high-order} perturbation by the gauge fluctuations  can result in such an anomaly. Instead, we first check the Landau diamagnetic susceptibility  {at a} high temperature limit, and in this case we have  
	\begin{eqnarray}
		\chi_f
		&\approx&\frac{g^2}{m_f^2T}(1-x),
	\end{eqnarray}
	and
	\begin{eqnarray}
		\chi_h
		&\approx &\frac{g^2T}{4m_h^2(2\pi)^2 N}\int d^2p\frac{1}{\xi_p\xi_{p+q}}
		\equiv \frac{g^2T}{T^4_0}.
	\end{eqnarray}
	Using the Hall resistivity of the free spinon and holon,  the Hall resistivity (\ref{HallR}) in the high temperature limit is approximated by 
	\begin{eqnarray}
		R_H\approx-\frac{1}{1-x}+ \frac1{x(1-x+\frac{T^2m^2_f}{T_0^4})}.\label{Eq:HallR_HT}
	\end{eqnarray}
	In the high temperature limit,  $R_H$ decreases as $T$  {increases}.  This implies that for the $t$-$J$ model, there is indeed a temperature interval where the Hall coefficient anomaly is consistent with  the experimental results for cuprates. 
	
	From the above analytical estimates, we see that the Hall resistivity increases at low  {temperatures} while  {decreasing} at high  {temperatures} as $T$  {increases}.

 \begin{figure}
		\vspace{3mm}
		\begin{minipage}{0.46\textwidth}
			\centerline{
				\includegraphics[width=\textwidth]{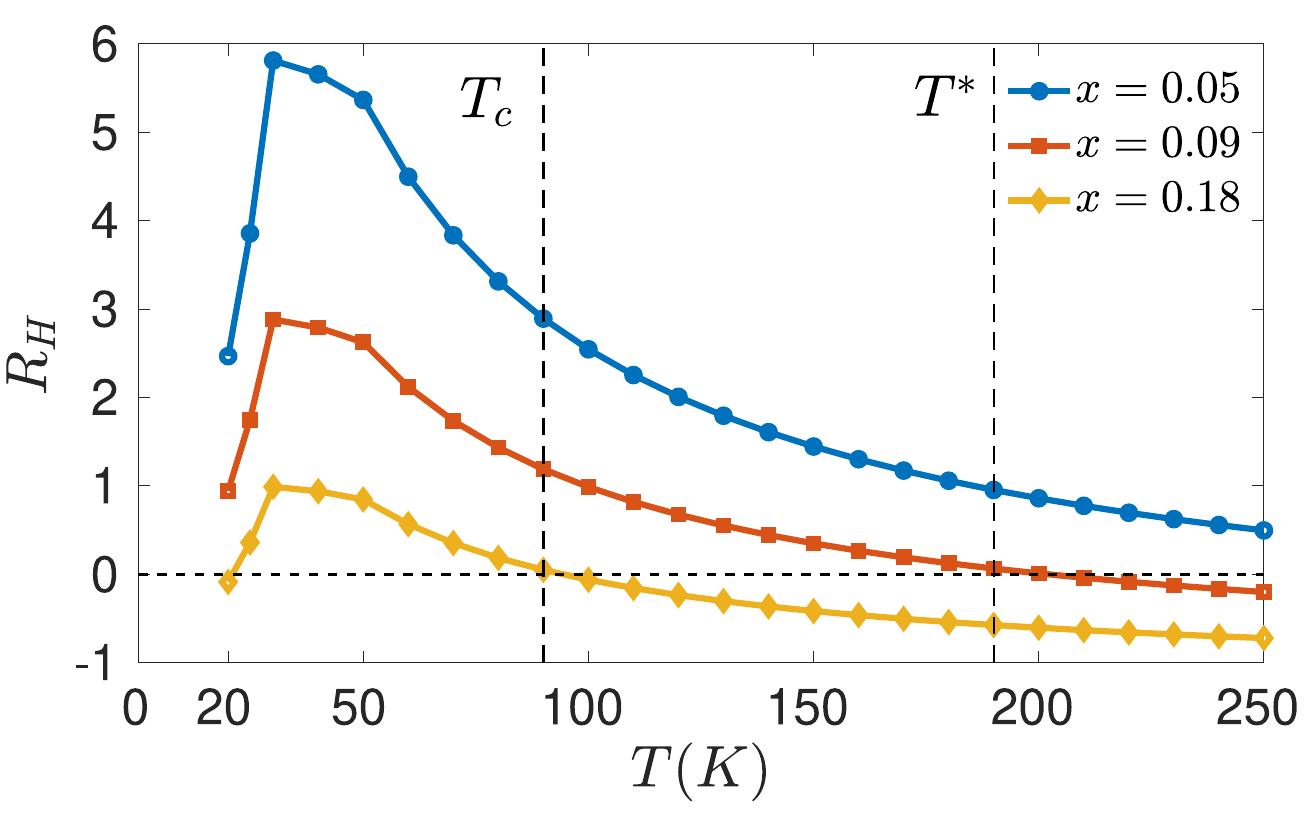}
			}	
		\end{minipage}
		\caption{(color online)  {Dependence of temperature and dopant concentration  {on} Hall resistivity. We choose ${\bf q}=(0.01,0.01)$ in Eqs.~\eqref{chif} and~\eqref{chih}. }}
		\label{fig10}	
\end{figure}

	
	In fact, with Eq.~\eqref{HallR}, we can calculate the Hall resistivity by numerically computing $\chi_{f,h}$ in Eqs.~\eqref{chif} and~\eqref{chih}. 
  In the long wavelength limit, numerical results qualitatively agree with  {the} above estimations {(See Fig.~\ref{fig10})}. We see that the numerical results are even better than the analytical estimations,  {compared} with the experiments~\cite{Fu,Hall1,Hall2}.  The increasing region of the Hall resistivity as $T$  {increases} is at  {a} very low temperature, in which the system is not in the strange metal phase. When $T>T^*$,  {where $T^\ast$ is the temperature at which strange metal appears},  the Hall resistivity monotonously decreases as $T$ raises for a given dopant concentration $x$.   On the other hand,  in  very low underdoping (say, $x=0.05$), the Hall coefficient is positive. This means that the observed Fermi surface is  {hole-type}. As $x$ increases (say $x=0.09$),  the Hall resistivity changes sign in some $T>T^*$. This implies that the observed Fermi surface changes from the hole type to  {the} electron type.  Close to the optimal doping (e.g., $x=0.18$), the Hall resistivity is always negative when $T>T^*$ and the Fermi surface is the electron type. This explains the puzzle of the Fermi surface that changes from the `small' one to the  `large' one as $x$ raises.

	\section{Conclusions and discussions}\label{sectionVI}
	
      In conclusion, we have formulated a  {consistent} $U(1)$ gauge theory 
      with   constraints on the local numbers and currents of the spinon and holon 
      within the slave boson representation of the $t$-$J$ model.  After considering further constraints on the Lagrange multipliers of the number and current constraints, the gauge fluctuations are dynamical. 
      The BRST symmetry plays an important role in the construction of the theory. Especially, in the ordered phases, the local number and current constraints are the {second-class} ones, and there  {is} no general method to quantize this system so that the gauge fixing conditions and the constraints are consistent.  However, we managed to develop a consistent example in this work by  {using} BRST quantization. We then focused on the strange metal phase, where the holons do not condense and the spinons are not paired.  We argued that in the uRVB state, the coupling constant of the gauge field is small, and then our model can be perturbatively solved.  Our gauge  theory method depends on the choice of mean field, and in principle, a renormalization group study would tell us when the mean field approximation becomes unreliable and new physics emerges.

	We computed the electron momentum distribution and demonstrated its non-Fermi liquid character. We calculated the electron spectral function and found that our result can explain the ARPES data for cuprates. {In the traditional slave boson treatment of the mean field theory, the occupation number distribution is independent of the momentum, which is  unphysical. In our method, we found that the normal state of the $t$-$J$ model in the strange metal phase  {was} not only not a Fermi liquid as expected but also not a Luttinger liquid or a marginal Fermi liquid because the Luttinger theorem was violated. Experimentally, this new physics could be confirmed by checking the occupation number distribution,  {which} can be obtained by summing over frequency in the ARPES data. This gives a new starting point to understand the strongly correlated physical phenomena such as various anomalous properties in the normal state of  {high-temperature} superconductivity, especially for the cuprates.} For the transport properties, we recovered the $T$ linear resistance, which is the main result of the previous gauge theory~\cite{NLPRL}. On the other hand, we found that, in the high temperature limit, although the resistivity is not linear in $T$, the Hall resistivity decreases as $T$ increases. This is the Hall coefficient anomaly, which was confirmed by the numerical calculation of the Hall resistivity in our theory. Furthermore, we revealed that the `large' Fermi surface crosses to the `small' Fermi surface as the dopant concentration and temperature  {vary}. 

 {Our theory is BRST invariant, so the results should not depend on the gauge fixing parameters. However, since in our calculations the RPA approximation was used, the results become gauge dependent. To restore  gauge invariance, other corrections which are of the same level of approximation as the RPA should be included. We can use the Ward-Takahashi identity as a guiding principle to develop gauge invariant approximations, which will be studied elsewhere.}

  The thermodynamics of the uRVB  was studied before, and the overestimation of the mean field entropy and the underestimation of the mean field free-energy loss were found in comparison with the exact results in the high temperature limit~\cite{EOver}. The gauge fluctuations may improve the free energy and entropy considerably. However, we do not study the thermodynamics with our BRST quantized gauge theory. This will be left  {for} further work.
  
 
 We only focused on the strange metal phase, but other phases are also important. Our previous work shows that the gauge fluctuations may improve the critical temperatures of the pseudogap and superconducting  phases~\cite{LLY}.
	We hope that these results can be recovered by using the perturbation framework developed in this article. Especially, the physics of the pseudogap state has been extensively studied recently, see, for example,~\cite{PS1,PS2}. 
 We expect to study the physical properties of this  phase more quantitatively to compare with experimental data  and the results from other theories.  
 {With proper modification, our theory can be applied to heavy fermion systems and may lead to a better understanding of the rich physical properties of these materials.} Our constrained gauge theory model in the slave boson (or fermion) representation may apply to other strongly correlated systems, including the  {topologically} non-trivial ones for which the BRST cohomology determines the topological structure of the quantum state space.  
	
	Finally, the BRST formalism developed in this work  {focuses} on the continuum theory around the $\Gamma$-point. If we want to solve the constrained problem around other  {highly} symmetric points, we need to obtain the continuum limit of the theory around these points and study the corresponding BRST formalism. We  {can} then   {perform} the perturbation calculations away from the $\Gamma$-point. 
	\\
	
	\noindent{\bf Acknowledgement} The authors thank Yan Chen, Jianhui Dai, Jianxin Li, Qian Niu, Ziqiang Wang, Congjun Wu, and Yong-Shi Wu  {for the useful discussions}.  We are  {particularly} grateful  {for} the insightful comments from Qianghua Wang on various physical problems. He pointed out that the vanishing counterflow current of the spinon and the holon is not  {a} first-class constraint. This enlightens us to study the BRST quantization with the second-class constraints in the mean field theory. This work is supported by the National Natural Science Foundation of China with Grant No.~12174067 (XL and YY) and Grant No.~12204329 (LL).

	\appendix
	
	\section{Constraints to the Lagrange multiplier}\label{ABRST}
	
	To understand the gauge theory with Dirac's first-class constraint,  we start  {with} the slave boson Hamiltonian  
	\begin{eqnarray}
		H_\lambda&=&H_0-\sum_i\lambda_iG_i=-\frac{J}4 \sum_{\langle ij\rangle}[|\gamma^f_{ij}|^2+|\Delta^f_{ij}|^2\nonumber\\
		&-&(\gamma^{f\dag}_{ij}\sum_\sigma f^\dag_{i\sigma} f_{j\sigma}+h.c.)]\nonumber\\
		&-&\sum_{\langle ij\rangle}\frac{J}4[\Delta^f_{ij}(f_{i\uparrow}^\dag f_{j\downarrow}^\dag-f_{i\downarrow}^\dag f_{j\uparrow}^\dag)+h.c.]\nonumber\\
		&+&t\sum_{\langle ij\rangle}h_ih^\dag_jf^\dag_{i\sigma}f_{j\sigma}-\sum_i\lambda_iG_i.\label{t-JH}
	\end{eqnarray}
	In Schr\"odinger's picture, $H$, $\lambda_i$ and $G_i$ are all time-independent.  Because $[H_\lambda,\lambda_i]=0$,  $\lambda_i$ does not evolve  {with} time $t$. Notice that here we use the real time $t$ but not the imaginary time $i\tau$. 
	
	\subsection{Additional constraint }
	
	To see why the additional constraint  {on} $\lambda_i$ is needed, we  {turn} to Heisenberg's picture. All operators and fields $\Phi_i$ become time-dependent,  $\Phi_i(t)=e^{iH_\lambda t}\Phi_ie^{-iH_\lambda t}$ except  {for} $\lambda_i$ since $[H_\lambda,\lambda_i]=0$, where  $ \Phi_i$  {represents} the 'matter' fields (holon, doublon, and spinon). 
      {To reveal the gauge structure of the problem and properly quantize it, we first  {promote} the Lagrange multiplier $\lambda_i$ to {a} dynamical variable.} 
 Thus, one has to add an additional constraint in order to be consistent with  {$\lambda_i(t)$ having no time} evolution, namely,
	\begin{eqnarray}
\partial_t\lambda_i(t)\equiv\dot\lambda_i(t)=0.
	\end{eqnarray}
	By introducing a new Lagrange multiplier $\pi_{\lambda_ i}(t)$ to force $\dot \lambda_i(t)=0$, the Lagrangian is then given by 
	\begin{eqnarray}
		L_\lambda&=&\sum_i\pi_{\lambda_i}(t)\dot\lambda_i(t)+\sum_{i\sigma}f^\dag_{i\sigma}(i\partial_t+\lambda_i(t)) f_{i\sigma}\nonumber\\
&+&\sum_i(h^\dag_i(i\partial_t+\lambda_i(t))h_i+\sum_id^\dag_i(i\partial_t+\lambda_i(t))d_i\nonumber\\
		&-&\sum_i\lambda_i(t)-H_0. \label{ALa}
	\end{eqnarray}
	
	\subsection{Classical field theory understanding}
	
	We may understand the relation between the Hamiltonian (\ref{t-JH}) and the Lagrangian (\ref{ALa}) from the point  {of} view of  classical field theory.  According to (\ref{ALa}), $\pi_{\lambda i}(t)=\frac{\delta L}{\delta \dot\lambda_i(t)}$, i.e., $\pi_{\lambda i}(t)$ is the canonical conjugate field of $\lambda_i(t)$.  Therefore, according to  classical mechanics, the Lagrangian for the Hamiltonian (\ref{t-JH}) reads
	\begin{eqnarray}
		L_\lambda=\sum_i (\pi_{\lambda i} \dot \lambda_i+\Pi_{\Phi_i} \dot \Phi_i)-H_\lambda, \label{ALaa}
	\end{eqnarray}
	where $\Pi_{\Phi_i}$  are the canonical conjugate fields of $ \Phi_i$, which  stand for the holon and spinon fields. The Lagrangian (\ref{ALaa}) is exactly the same as (\ref{ALa}).
	
	\subsection{Gauge symmetry}
	
	We now explain the reason  {for adding} the constraint $\dot\lambda_i(t)=0$ from the gauge symmetry point of view. 
	In  {the} literature,  instead of (\ref{ALa}), the following Lagrangian is considered  \cite{B,zou}  
	\begin{eqnarray}
		L_{GI}&=&\sum_{i\sigma}f^\dag_{i\sigma}(i\partial_t+\lambda_i(t)) f_{i\sigma}+\sum_i(h^\dag_i(i\partial_t+\lambda_i(t))h_i
		\nonumber\\
&+&\sum_id^\dag_i(i\partial_t+\lambda_i(t))d_i-H_\lambda \label{LGI}.
	\end{eqnarray}
	It  {is} known that the electron operator $c^\dag_{i\sigma}= f^\dag_{i\sigma}h_i +\sigma f_{i,-\sigma}d_i^\dag$ is gauge invariant under  $(h_i,d_i, f_{i\sigma})\to e^{-i\theta_i} (h_i,d_i, f_{i\sigma})$. $L_{GI}$ is invariant under this gauge transformation,  {accompanied by} $\lambda_i(t)\to \lambda_i(t)-\dot\theta_i$, i.e., $\lambda_i(t)$  plays a role of a scalar gauge potential.  There are redundant gauge degrees of freedom in the path integral 
	\begin{eqnarray}  
		W'&=&\int \prod_{i,t} d\Phi_i^\dag(t) d\Phi_i(t) d\lambda_i(t)e^{i\int dt L_{GI} }.\label{W'}
	\end{eqnarray}
	One way to remove the redundant gauge degrees of freedom is  {by} taking the gauge fixing $\dot\lambda_i(t)=0$, namely, replacing $L_{GI}$  {with} the Lagrangian (\ref{ALa}), the path integral reads
	\begin{eqnarray}
		W=\int \prod_{i,t} d\Phi_i^\dag(t) d\Phi_i(t) d\pi_{\lambda_i}(t) d\lambda_i(t)e^{i\int dt L_\lambda }. \label{W}
	\end{eqnarray}
	For the gauge theory, we can make a gauge transformation 
	\begin{eqnarray}
		\dot\lambda_i\to\dot\lambda_i+\xi \pi_{\lambda i}, \label{lambdaG}
	\end{eqnarray}
	for Eq. (\ref{W}), where $\xi$ is an arbitrary constant, {and then the $\dot{\lambda}_i\pi_{\lambda_i}$ term changes to}
 \begin{eqnarray}
     \dot{\lambda}_i\pi_{\lambda_i}+\xi \pi^2_{\lambda_i}.
 \end{eqnarray}
 Integrating away  {the} $\pi_{\lambda i}$ field, the path integral becomes
	\begin{eqnarray}
		W\propto \int \prod_{i,t} d\Phi_i^\dag(t) d\Phi_i(t) d\lambda_i(t)  e^{i\int dt L_{\rm eff}},\label{W1}
	\end{eqnarray}
	where 
	\begin{eqnarray}
		L_{\rm eff}=L_{GI}-\frac{1}{2\xi}\sum_i\dot\lambda_i^2(t).
	\end{eqnarray}
	This is a correct gauge fixing Lagrangian of the Abelian gauge theory, but Eq. (\ref{W1}) is not gauge invariant.
	In order to resolve this paradox,  we recall  {the} Faddeev-Popov quantization of the gauge theory. We insert 1 into the gauge invariant (\ref{W'}) to fix the redundant gauge degrees of freedom in terms of
	\begin{eqnarray}
		1=\int \prod_{i,t}d\theta_{i,t} \delta(\dot\lambda_i(t)){\rm det}(\frac{\delta\dot\lambda_i(t)}{\delta\theta_j(t')}),
	\end{eqnarray}
	and finally \cite{Peskin},
	\begin{eqnarray}
		1\cdot W'&=&N(\xi)\int \prod_{i,t} d\Phi_i^\dag(t) d\Phi_i(t) d\lambda_i(t){\rm det }(\partial^2_t)\nonumber\\
		&\times&\exp\{i\int dt L_{\rm eff}\},\label{GFL}
	\end{eqnarray}
	where $N(\xi)$ is an unimportant infinity constant. The path integral (\ref{GFL}) is gauge invariant.  Comparing (\ref{GFL}) and (\ref{W1}), they differ from a factor ${\rm det }(\partial^2_t)$ after dropping $N(\xi)$.  {In} the present case, this determinant does not contain any fields and is a constant. This means that  (\ref{W1}) is equivalent to (\ref{GFL}). Therefore, up to a constant determinant, (\ref{W1})  is gauge invariant.  However, for a non-Abelian gauge theory, the determinant in general is dependent on the gauge field and cannot be dropped. This is why Faddeev-Popov ghost fields are introduced.  
	
	At finite temperature, we replace $t\to i\tau$ and finally obtain the effective Lagrangian (\ref{ETL}) in the main text. 
	
	\section{General form of quadratic gauge fixing conditions with BRST invariance}\label{general}

  {In this appendix, we {will demonstrate} how to {identify} consistent gauge fixing conditions. Our guiding principle  {will be} the BRST invariance.} 

  {We first write down possible  gauge fixing conditions with the ghost term.} 
Since our theory is not Lorentz invariant, the quadratic gauge fixing condition with the ghost term in the Lagrangian can be generalized from the Lorenz gauge to the following general form:
	\begin{eqnarray}\label{Eq:LGFgh}
		\mathcal{L}_{GF+gh}&=&\frac{A}{2}(\partial_\tau\delta\lambda)^2+B\sum_b \partial_\tau \delta a_b\partial_b\delta \lambda
		+\frac{C}{2}(\sum_b \partial_b \delta a_b)^2\nonumber\\
		&&+\frac{D}{2}\sum_b (\partial_\tau \delta a_b)^2+\frac{E}{2}\sum_b(\partial_b \delta\lambda)^2+\bar{u}K u,
		%
	\end{eqnarray}
	where the coefficients $A$, $B$, $C$, $D$ and $E$ are  {constants}, and $K(\partial_\tau,\partial_b)$ is an operator describing the dynamics of the ghost field. The parameters $A$, $B$, $C$, $D$ and $E$ and the operator $K(\partial_\tau,\partial_b)$  should be determined by the BRST invariance. We assume that $K$ is independent of the gauge fields, i.e., the coupling between ghost and gauge fields is absent. We will see that either $B=0$ or $E=0$ is allowed.
	In the main text, we take
	\begin{eqnarray}\label{Eq:GaugeParameters}
		A=-\zeta,\quad B=0,\quad C=-\frac{1}{\xi}, \quad D=-1, \quad E=-\frac{\zeta}{\xi},\nonumber\\
	\end{eqnarray}
 {to simplify perturbative calculations. }
	
	\subsection{BRST invariance}
 
 
 {We denote the infinitesimal BRST transformation by $\delta_\epsilon$, where $\epsilon$ is an infinitesimal {Grassmann} constant. It is convenient to write $\delta_\theta$ as $\delta_\epsilon\equiv \epsilon s$ with $s$ being a fermion operator~\cite{Weinberg}. Then the infinitesimal BRST transformation for the gauge fields $\delta\lambda$ and $\delta a_b$ as well as  {the} ghost field $u$ is}
	\begin{eqnarray}
		\delta_\epsilon \delta\lambda=\epsilon\partial_\tau u=\epsilon s \delta\lambda,~\delta_\epsilon \delta a_b=\epsilon \partial_b u=\epsilon s \delta a_b,~\delta_\epsilon u=0=\epsilon s u,\nonumber\\
	\end{eqnarray}
	and since the operator $K$ is not known yet, the transformation rule of  {the} anti-ghost $\bar{u}$ is to be determined. 
  {The effect of the BRST transformation, acting on the matter and  {matter}-gauge coupling terms, is the same as that of the gauge transformation. Therefore the matter and matter-gauge coupling sector is invariant under the transformation. The changes come from the gauge fixing and ghost parts, Eq.~\eqref{Eq:LGFgh}.} Under the BRST transformation, the Lagrangian density $\mathcal{L}$ changes as
	\begin{eqnarray}\label{Eq:sL}
		s\mathcal{L}&=&A\partial_\tau\delta\lambda \partial^2_\tau u+B\sum_b\partial_\tau\partial_b u\partial_b \delta\lambda\nonumber\\
		&&+B\sum_b\partial_\tau\delta a_b\partial_b\partial_\tau u
		+C\sum_b\partial_b\delta a_b\sum_c \partial_c \partial_c u\nonumber\\
		&&+D\sum_b \partial_\tau\delta a_b \partial_\tau \partial_b u+E\sum_b \partial_b\delta\lambda\partial_b\partial_\tau u+s\bar{u}K u,\nonumber\\
		&=&A\partial_\tau\delta\lambda \partial^2_\tau u
		+(B+E)\sum_b\partial_\tau\delta\lambda \partial_b\partial_b u \nonumber\\
		&&+(B+D)\sum_b\partial_b \delta a_b\partial^2_\tau u
		+C\sum_b\partial_b\delta a_b\sum_c \partial_c \partial_c u\nonumber\\
		&&+s\bar{u}Ku+\partial_\mu K^\mu,\nonumber\\
		&=&[A\partial_\tau\delta\lambda +(B+D)\sum_b\partial_b \delta a_b ]\partial^2_\tau u\nonumber\\
		&&+[(B+E)\partial_\tau\delta\lambda+C\sum_b\partial_b \delta a_b]\sum_b\partial_b^2u\nonumber\\
		&&+s\bar{u}Ku+\partial_\mu K^\mu,
	\end{eqnarray}
	where
	\begin{eqnarray}
		\partial_\mu K^\mu &=&B\partial_\tau (\sum_b \partial_b u\partial_b \delta\lambda)-B\sum_b \partial_b(\partial_b u \partial_\tau \delta\lambda)\nonumber\\
		&&	+B\partial_\tau (\sum_b \delta a_b\partial_b\partial_\tau u)-B \sum_b\partial_b( \delta a_b\partial^2_\tau u)\nonumber\\
		&&+D\partial_\tau(\sum_b \delta a_b\partial_\tau\partial_b u)
		-D\sum_b \partial_b(\delta a_b\partial^2_\tau u)\nonumber\\
		&&+E\partial_\tau(\sum_b \partial_b\delta\lambda\partial_b u)
		-E\sum_b \partial_b(\partial_\tau\delta\lambda\partial_b u).\nonumber\\
	\end{eqnarray}
	The BRST invariance of the theory requires that the Lagrangian density is invariant up to a total derivative,   {and therefore the first three terms in Eq.~\eqref{Eq:sL} must vanish identically,}
	\begin{eqnarray}
		&&[A\partial_\tau\delta\lambda +(B+D)\sum_b\partial_b\delta a_b ]\partial^2_\tau u\nonumber\\
		&&+[(B+E)\partial_\tau\delta\lambda+C\sum_b\partial_b \delta a_b]\sum_b\partial_b^2u+s\bar{u}Ku=0,~~~~~~~~
	\end{eqnarray}
	which leads to{
		\begin{eqnarray}
			\frac{A}{B+E}&=&\frac{B+D}{C}\equiv \xi,~\mathrm{or}~\frac{B+E}{C}=\frac{A}{B+D}\equiv \zeta,\label{Eq:ABCDE_relation} \nonumber\\\\
			K&=&-C(\xi\partial^2_\tau +\sum_b\partial_b^2),\\
		s\bar{u}&=&\zeta\partial_\tau\delta\lambda +\sum_b\partial_b \delta a_b.
	\end{eqnarray} }
 {We thus find relations between the gauge fixing parameters and also determine the ghost Lagrangian and the BRST transformation for the antighost field.}

	 {For completeness, we write down the equations of motion for $\delta\lambda$, $\delta a_b$, and $u$}:
	\begin{eqnarray}
		A\partial^2_\tau\delta\lambda +B\sum_b\partial_\tau\partial_b \delta a_b+E\sum_b\partial_b^2\delta\lambda&=&-i g G, \\
		B\partial_\tau \partial_b\delta\lambda+C\partial_b(\sum_c\partial_c \delta a_c)+D\partial^2_\tau \delta a_b&=&J_b, \\
		(\xi \partial^2_\tau+\sum_b\partial_b^2)u&=&0,
	\end{eqnarray}
	where $G$ is the local constraint in the continuous limit defined in the main text.
	Note that if we take $\xi=\zeta$, then $ss\bar{u}=0$ due to the equation of motion of $u$.

	\subsection{The BRST charge}
 {The BRST symmetry is a global symmetry, thus according  {to} Noether's theorem, there is a conserved charge that {generates} the transformation. In this subsection we calculate the BRST charge $Q$.} Under the BRST transformation ($\epsilon$ is an anti- {commuting} constant), the action changes as
	\begin{eqnarray}
		\delta_\epsilon S&=&\int\mathrm{d}^3x \frac{\partial \mathcal{L}}{\partial \partial_\mu \Phi}\delta_\epsilon\partial_\mu \Phi+\frac{\partial \mathcal{L}}{\partial \Phi}\delta_\epsilon \Phi\nonumber\\
		&=&\int\mathrm{d}^4x \frac{\partial \mathcal{L}}{\partial \partial_\mu \Phi}\delta_\epsilon \partial_\mu \Phi+\partial_\mu\frac{\partial \mathcal{L}}{\partial \partial_\mu\Phi}\delta_\epsilon \Phi,\nonumber\\
		&=&\int\mathrm{d}^3x \partial_\mu(\frac{\partial \mathcal{L}}{\partial \partial_\mu \Phi}\delta_\epsilon  \Phi).
	\end{eqnarray}
 {To  {obtain} the second equation from the first one, we  {utilized} the equations of motion of the fields and  {conducted} integration by parts.} On the other hand [see Eq.~\eqref{Eq:sL}]
	\begin{eqnarray}
		\delta_\epsilon S=\epsilon\int\mathrm{d}^4x \partial_\mu K^\mu.
	\end{eqnarray}
  {Comparing the above expressions, we find that}
	\begin{eqnarray}
		\epsilon Q&=&\int\mathrm{d}^2x (\frac{\partial \mathcal{L}}{\partial \partial_\tau \Phi}\delta_\epsilon  \Phi-\epsilon K^\tau),\\
		&=&\epsilon\int \mathrm{d}^2x~ (igGu+A\partial_\tau\delta\lambda \partial_\tau u+B\sum_b \partial_b\delta\lambda\partial_b u\nonumber\\
		&&+D\sum_b\partial_\tau \delta a_b\partial_b u
		-B\sum_b \partial_b u\partial_b \lambda-B\sum_b \delta a_b\partial_b\partial_\tau u\nonumber\\
		&&-D\sum_b \delta a_b\partial_\tau\partial_b u-E\sum_b \partial_b\delta\lambda\partial_b u
		),\nonumber\\
		&=&\epsilon\int \mathrm{d}^2x~ (igGu+A\partial_\tau\delta\lambda \partial_\tau u+D\sum_b\partial_\tau \delta a_b\partial_b u\nonumber\\
		&&-B\sum_b a_b\partial_b\partial_\tau u-D\sum_b \delta a_b\partial_\tau\partial_b u-E\sum_b \partial_b\delta\lambda\partial_b u
		),\nonumber\\
	\end{eqnarray}
	i.e.,
	\begin{eqnarray}
		Q&=&\int \mathrm{d}^2x~ (igGu+A\partial_\tau\delta\lambda \partial_\tau u+D\sum_b\partial_\tau \delta a_b\partial_b u\nonumber\\
		&&-B\sum_b \delta a_b\partial_b\partial_\tau u-D\sum_b a_b\partial_\tau\partial_b u-E\sum_b \partial_b\delta\lambda\partial_b u),\nonumber\\
		&=&\int \mathrm{d}^2x~ (igG +E\sum_b\partial_b^2\delta\lambda-D\sum_b\partial_\tau\partial_b \delta a_b )u\nonumber\\
		&&+[A\partial_\tau\delta\lambda +(B+D)\sum_b\partial_b \delta a_b]\partial_\tau u.
	\end{eqnarray}
	The BRST charge can also be obtained by calculating  $\delta_{\epsilon(\tau,\mathbf{r})}S$, which gives the same result.
 {Physical states must be annihilated by the BRST charge, and thus we shall have the constraints~\cite{BRST_primer},}
	\begin{eqnarray}
		&&G=0,\\
		&&E\sum_b\partial_b^2\delta\lambda-D\sum_b\partial_\tau\partial_b \delta a_b =0,\label{Eq:GFC1}\\
		&&A\partial_\tau\delta \lambda +(B+D)\sum_b\partial_b \delta a_b=0.\label{Eq:GFC2}
	\end{eqnarray}
 {Substituting the parameters~Eq.~\eqref{Eq:GaugeParameters} into Eq.~\eqref{Eq:GFC1} and Eq.~\eqref{Eq:GFC2}, we get the gauge fixing conditions Eqs.~\eqref{GF1} and~\eqref{LG} in the main text.}

	\subsection{Gauge field Green's function}
	 {After gauge fixing, the gauge field Green's function can be determined.}	The inverse  {of} Mastubara Green's function for the gauge field is
	\begin{eqnarray}
		&&	\mathcal{D}^{(0)-1}(\mathbf{k},i\nu_n)=\nonumber\\
		&&	\left[
		\begin{array}{cccc}
			A\nu^2_n+E\mathbf{k}^2 & B\nu_n k_x & B\nu_n k_y \\
			B\nu_n k_x & D\nu^2_n+Ck^2_x & Ck_x k_y\\
			B\nu_n k_y & C k_x k_y & D\nu^2_n+Ck^2_y
		\end{array}
		\right],
	\end{eqnarray}
	and
	\begin{eqnarray}
		&&\det \mathcal{D}^{(0)-1}=\nonumber\\
		&&D\nu^2_n[AD\nu^4_n+(AC+DE-B^2){\bf k}^2\nu^2_n+CE {\bf k}^4].\nonumber\\
	\end{eqnarray}
	When $D\to 0$, ${\cal D}_{\mu\nu}^{(0)}$ is not well-defined, which means that  $\sum_b\partial_b^2 \delta\lambda$ cannot be zero in order to  {perform} the well-defined gauge fixings.
	
	For $B=0$, the temporal and spatial components are decoupled, and Eq.~\eqref{Eq:ABCDE_relation} becomes
	\begin{eqnarray}
		\frac{A}{E}&=&\frac{D}{C}\equiv \xi,~\mathrm{or}~\frac{E}{C}=\frac{A}{D}\equiv \zeta, 
	\end{eqnarray}
	and 
	the Green's function is simplified as
	\begin{eqnarray}
		&&	\mathcal{D}^{(0)00}(\mathbf{k},i\nu_n)=\frac{1/A}{\nu^2_n+\mathbf{k}^2/\xi},\\
		&&	\mathcal{D}^{(0)ij}(i\nu_n,\mathbf{k})=\nonumber\\&&\frac{1/D}{\nu^2_n(\nu^2_n+k^2/\xi)}
		\left[
		\begin{array}{ccc}
			\nu^2_n+(k^2-k^2_x)/\xi & -k_x k_y/\xi \\
			-k_x k_y/\xi & \nu^2_n+(k^2-k^2_y)/\xi 
		\end{array}
		\right].\nonumber\\
	\end{eqnarray}

	\section{Anomalous Green's functions} \label{AAn}
	
\begin{figure}
		\vspace{3mm}
		\includegraphics[width=0.3\textwidth]{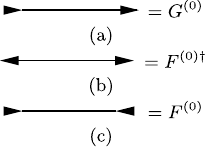}
		\caption{
			{ The Feynman diagrams for the anomalous Green's functions of (a) $G^{(0)}$, (b) $F^{(0)\dag}$, and (c) $F^{(0)}$. }
			\label{app-fig10}	}
	\end{figure}

	In the fermion paired phases, the effective Lagrangian  {is} rewritten  {using} the Nambu representation, and  the anomalous Green's functions are defined by
	\begin{eqnarray}
		G({\bf k},\tau-\tau')&=&-\langle T_\tau \psi_{k,\sigma}(\tau) \psi^\dag_{k,\sigma} (\tau') \rangle, \nonumber\\
		F({\bf k},\tau-\tau')&=&\langle T_\tau \psi_{-k,\downarrow}(\tau) \psi_{k,\uparrow} (\tau') \rangle,\nonumber\\
		F^\dag({\bf k},\tau-\tau')&=&\langle T_\tau \psi^\dag_{k,\uparrow}(\tau) \psi^\dag_{-k,\downarrow} (\tau')\rangle.
	\end{eqnarray}
	And then 
	\begin{eqnarray}
		{ G}^{(0)}({\bf k},i\omega_n)&=&\frac{u_p^2}{i\omega_n-E_k}+\frac{v_k^2}{i\omega_n+E_k}, \nonumber\\
		{ F}^{(0)}({\bf k},i\omega_n)&=&{ F}^{(0)\dag}({\bf k},i\omega_n)\nonumber\\
		&=&-u_pv_p(\frac{1}{i\omega_n-E_k}-\frac{1}{i\omega_n+E_k}),
	\end{eqnarray}
	where $u_k, v_k$ and $E_k$ follow the standard BCS notion. The Feynman diagrams are shown in Fig.~\ref{app-fig10}.

	\section{Feynman's rules and Dyson's equations}\label{feynmanRule}

	In this appendix, we  {provide} the Feynman's rules and Dyson's equations. The Feynman's rules are
	
	(1) For each line, associate a corresponding free one-particle Green's  {function}  where
	$\nu_n=2n\pi/\beta$ for bosons and $\omega_n=(2n+1)\pi n/\beta$ for fermions. 
	
	(2) Conservation of energy and momentum at each vertex.
	
	(3) Sum over internal degrees of freedom: momentum, energy, and spin, including the momentum and energy associated with loop diagrams.
	
	(4) Finally, multiply each diagram by the factor 
	\begin{equation}
		\frac{(-1)^{m+F}}{(2\pi)^{3}},
	\end{equation}
	where $F$ is the number of closed fermion loops. For the fermion self-energy, $m$ is the number of internal gauge field lines. For the boson self-energy (or vacuum polarization), $m$ is one-half of the number of vertices. 
 
	For the spinon $\psi_\sigma$ and holon $h$, the full one-particle Green's functions are given by Dyson equations
	\begin{equation}
		{\cal G}({\bf k},i\omega_n)=\frac{{\cal G}^{(0)}({\bf k},i\omega_n)}{1-{\cal G}^{(0)}({\bf k},i\omega_n)\Sigma({\bf k},i\omega_n)},\label{mdyson}
	\end{equation}
	where $\Sigma$ is the self-energy.  Under the random phase approximation (RPA), the self-energies are replaced by $\Sigma^{(0)}$ in (\ref{mdyson})  with $\Sigma^{(0)}$ being the bubbles in the one-loop diagrams.

	For the gauge fields, the full one-particle Green's function is given by Dyson's equation
	\begin{equation}
		{\cal D}({\bf q},i\nu_n)=\frac{{\cal D}^{(0)}({\bf q},i\nu_n)}{1-{\cal D}^{(0)}({\bf q},i\nu_n)\Pi({\bf q},i\nu_n)},\label{gdyson}
	\end{equation}
	where $\Pi$ is the vacuum polarization.

	\section{Momentum distribution}\label{momentum_dis}
 Here we present calculations for the momentum distribution function. 
		The $g^2$ order correction to the source vertex comes from the  {fifth} line of Fig.~\ref{fig4}. The former is still a constant, while the latter is dependent on ${\bf k}$ and given by
	\begin{eqnarray}
		&&Z_{eff}^{-1}\sum_{a,b,{\bf q,q'},\nu_m,\nu'_l}\int \prod d\Phi e^{-S_0}\sum_{\sigma'{k', k'',p,p' }}\nonumber\\
		&&\frac{g^2}{2m_f m_h} f^\dag_{\sigma' {k'}}\delta a_a(-p)f_{\sigma' {k'+p}}(2k'_a+p_a) 
		\nonumber\\
		&\times&h^\dag_{ k''+p'}\delta a_b^\dag(-p')h_{  k''}(2k''_b+p'_b) 
		\nonumber\\
		&\times&f^\dag_{\sigma {\bf k+q},i\nu_m}f_{\sigma,{\bf k+q'},i\nu'_l}h_{q}h^\dag_{q'},\label{g2}
	\end{eqnarray}
	where $k'$ stands for $(i\omega_n, {\bf k'})$ and so on. At zero temperature, we can separate the ${\bf k}$-dependent part in Eq.~\eqref{g2}, which is given by
	\begin{eqnarray}
		&&\frac{g^2}{m_fm_h}\sum_{a,b,{\bf q,q'},\omega_1,\omega_2}(q_a+q'_a) (2k_b)\nonumber\\
		&&{\rm Re}{\cal D}^{(0)}_{ab}(-{\bf (q-q')},\omega_1-\omega_2)\nonumber\\
		&&\Theta(-\omega_1) \Theta(-\omega_2) A^{(0)}_f({\bf q'},\omega_1) A^{(0)}_f({\bf q},\omega_2).~~~\label{g2o}
	\end{eqnarray}
	Due to the reflection symmetry, Eq.~\eqref{g2o} vanishes. Hence, the $g^2$ order contribution to the momentum distribution is also a constant.

	Let us check the $g^4$ order contribution. Besides the linearly $\bf k$-dependent diagrams, which eventually become zero due to reflection symmetry, the diagrams in Fig.~\ref{fig5} provide the nonzero quadratic ${ k_ak_b}$-dependent contributions to the electron momentum distribution. Namely, corresponding to {Fig.~\ref{fig5}(a) and~\ref{fig5}(b)}, the contributions to the electron Green's functions are given by 
	\begin{widetext}
		\begin{eqnarray}
			{\cal G}_{e\sigma}^{4A}({\bf k},i\omega_n)&=&-\sum_{m_1,m_2,m_3;\bf q_1,q_2,q_3 }{\cal G}^{(0)}_{f\sigma}({\bf k+q_1},i\omega_n+i\nu_{m_1}){\cal G}^{(0)}_{f\sigma}({\bf k+q_2},i\omega_n+i\nu_{m_2}){\cal G}^{(0)}_{f\sigma}({\bf k+q_3},i\omega+i\nu_{m_3})\nonumber\\
			&&\times {\cal G}^{(0)}_h({\bf q_1},i\nu_{m_1}){\cal G}^{(0)}_h({\bf q_2},i\nu_{m_2}){\cal G}^{(0)}_h({\bf q_3},i\nu_{m_3})\sum_{a,b,c,d}{\cal D}^{(0)}_{ab}({\bf q_1-q_2},i\nu_{m_1}-i\nu_{n_2})
			\nonumber\\
			&&\times{\cal D}^{(0)}_{cd}({\bf q_3-q_2},i\nu_{m_3}-i\nu_{m_2})(q_{1a}+q_{2a})(2k_b+q_{1b}+q_{2b})(q_{2c}+q_{3c})(2k_d+q_{2d}+q_{3d})\frac{g^4}{m_f^2m_h^2},\\
			{\cal G}_{e\sigma}^{4B}({\bf k},i\omega_n)&=&-\sum_{m_1,m_2,m_3;\bf q_1,q_2,q_3 }{\cal G}^{(0)}_{f\sigma}({\bf k+q_1},i\omega_n+i\nu_{m_1}){\cal G}^{(0)}_{f\sigma}({\bf k+q_1-q_2+q_3},i\omega_n+i\nu_{m_1}-i\nu_{m_2}+i\nu_{m_3})\nonumber\\
			&&\times{\cal G}^{(0)}_{f\sigma}({\bf k+q_3},i\omega_n+i\nu_{m_3}) {\cal G}^{(0)}_h({\bf q_1},i\nu_{m_1}){\cal G}^{(0)}_h({\bf q_2},i\nu_{m_2}){\cal G}^{(0)}_h({\bf q_3},i\nu_{m_3})\nonumber\\
			&&\times \sum_{a,b,c,d}{\cal D}^{(0)}_{ab}({\bf q_1-q_2},i\nu_{m_1}-i\nu_{m_2}){\cal D}^{(0)}_{cd}({\bf q_3-q_2},i\nu_{m_3}-i\nu_{m_2})\nonumber\\
			&&\times(q_{1a}+q_{2a})(2k_b+2q_{1b}-q_{2b}+q_{3b})(q_{2c}+q_{3c})(2k_d+q_{1d}-q_{2d}+2q_{3d})\frac{g^4}{m_f^2m_h^2}.
		\end{eqnarray}
	\end{widetext}
	
	At zero temperature, the contributions to the electron momentum distribution  {are as follows:}
	\begin{widetext}		
		\begin{eqnarray}
			n_{e\sigma\bf k}^{4A}&\approx&-\sum_{a,b,,c,d,\nu_1,\nu_2,\nu_3, \bf q_1,q_2,q_3 }\Theta(-\nu_1)\Theta(-\nu_2)\Theta(-\nu_3)A^{(0)}_{f\sigma}({\bf q_1},\omega+\nu_1)A^{(0)}_{f\sigma}({\bf q_2},\omega+\nu_2)A^{(0)}_{f\sigma}({\bf q_3},\omega+\nu_3))\nonumber\\
			&&\times D^{(0)}_{ab}({\bf q_1-q_2},\nu_1-\nu_2)D^{(0)}_{cd}({\bf q_3-q_2},\nu_3-\nu_2)(-2k_a)(q_{1b}+q_{2b})(-2k_c)(q_{2d}+q_{3d})\frac{g^4}{m_f^2m_h^2}\nonumber\\
			&\equiv&-\sum_{b,c}{\cal C}^{4A}_{bc}k_bk_c,\\
			n_{e\sigma\bf k}^{4B}&\approx&-\sum_{a,b,c,d,\nu_1,\nu_2,\nu_3, \bf q_1,q_2,q_3 }\Theta(-\nu_1)\Theta(-\nu_2)\Theta(-\nu_3)A^{(0)}_{f\sigma}({\bf q_1},\omega+\nu_1)A^{(0)}_{f\sigma}({\bf q_1-q_2+q_3},\omega+\nu_1-\nu_2+\nu_3))\nonumber\\
			&&\times A^{(0)}_{f\sigma}({\bf k+q_3},\omega+\nu_3) D^{(0)}_{ab}({\bf q_1-q_2},\nu_1-\nu_2)D^{(0)}_{cd}({\bf q_3-q_2},\nu_3-\nu_2)\nonumber\\
			&&\times(-2k_a)(2q_{1b}-q_{2b}+q_{3b})(-2k_c)(q_{1d}-q_{2d}+2q_{3d})\frac{g^4}{m_f^2m_h^2}\nonumber\\
			&\equiv&-\sum_{a,c}{\cal C}^{4B}_{ac}k_ak_c.
		\end{eqnarray}
	\end{widetext}
	Due to the rotational symmetry, we have 
	\begin{eqnarray}
		n^{(4)}_{e{\bf k},T=0}=2(n_{e\sigma\bf k}^{4A}+n_{e\sigma\bf k}^{4B})\equiv -\sum_{ab}C^{(4)}_{ab}k_ak_b=-C^{(4)}k^2,\nonumber
	\end{eqnarray}
	as $C^{(4)}_{ab}=C^{(4)}\delta_{ab}$.
	Other contributions to the $k^2$ terms come from the correction of ${\cal G}_{00}$ to $n^{(4)}_{e{\bf k},T=0}$, which is of the order $O(g^6)$ and higher. This corrects $C^{(4)}\to \tilde C^{(4)}$. Similarly, the $2n$- {loop} diagrams with the spinon-holon vertex  {and} $2n-1$ lines of the spatial gauge field Green's function do not contribute to the $g^{4n-2}$ order,  while such $2n+1$- {loop} diagrams contribute to $n_{e\bf k}$ with
	\begin{eqnarray}
		n^{(2n+2)}_{e{\bf k}, T=0}=(-1)^nC^{(2n+2)} k^{2n},
	\end{eqnarray}  
	and $C^{(2n+2)}$ is corrected to $\tilde C^{(2n+2)}$. The momentum distribution at $T=0$ is then given by
	\begin{eqnarray}
		n_{e{\bf k},T=0}=\sum_{n=0} (-1)^n\tilde C^{(2n+2)} k^{2n}.
	\end{eqnarray}
 
 	\section{Spinon and holon self-energies} \label{SHSelfEnergy}
  In this appendix, we present  expressions for spinon and holon self-energies.  We first integrate over the azimuth angle $\phi$, and the retarded self-energy of the spinon reads  
	\begin{widetext}
		\begin{eqnarray}
			\Sigma^{(0)}_{f}(\omega+i0^+,\mathbf{q})
			&=&g^2\sum_{s=\pm}\int\frac{kdk}{2\pi}~
			\frac{sn_B(s k)m_f}{2k^2q}\left[
			\frac{k^2 I^+_1(\omega_{s,f}) +kq I^+_2(\omega_{s,f})+q^2I^+_3(\omega_{s,f})/4}{m^2_f}+
			I^+_1(\omega_{s,f})\right]\nonumber\\
			&&-g^2\int\frac{kdk}{2\pi}~
			\frac{n_F(\xi_{f,{\bf k}})}{2kq}\left[
			J_1(\omega_f)+\frac{(k^2+q^2/4)J_1(\omega_f)+kqJ_2(\omega_f)}{m^2_f}
			\right]\nonumber\\
			&&+\frac{g^2}{m^2_f}\int\frac{kdk}{2\pi}~
			\frac{n_F(\xi_{f,{\bf k}})kq[J_1(\omega_f)-J_3(\omega_f)]}{8(\omega+i0^+-\xi_{f,{\bf k}})^2},
		\end{eqnarray}
	\end{widetext}
	where 
 \begin{eqnarray}
     \omega_{s,f}&=&\frac{m_f}{kq}(\omega+sk-\frac{k^2+q^2}{2m_f}+\mu_f),
\end{eqnarray}
and
 \begin{eqnarray}
     \omega_{f}&=&\frac{1}{2kq}((\omega-\xi_{f,\bf k})^2-k^2-q^2).
 \end{eqnarray}
The functions $I^\pm_{1,2,3}(x)$ are defined by
	\begin{eqnarray}
		&&I^{\pm}_{1}(x)
		=\frac{\mathrm{sgn}(x)\theta(|x|-1)}{\sqrt{x^2-1}} \mp i \frac{\theta(1-|x|)}{\sqrt{1-x^2}},\nonumber\\
		&&I^{\pm}_{2}(x)=-1+x I^{\pm}_1(x),\nonumber\\
		&&I^{\pm}_{3}(x)=-x+x^2 I^{\pm}_1(x).\nonumber 
	\end{eqnarray}
	And $J_{1,2,3}(\omega_f)$ are given by 
	\begin{eqnarray}
		J_{1,2,3}(\omega_f)=I_{1,2,3}^{{\rm sgn}(\omega-\xi_{f,\bf k})}(\omega_f).
	\end{eqnarray}
	
	Similarly, the retarded holon self-energy is given by  
	\begin{widetext}
		\begin{eqnarray}
			\Sigma^{(0)}_{h}(\omega+i0^+,\mathbf{q})
			&=&g^2\sum_{s=\pm}\int\frac{kdk}{2\pi}~
			\frac{sn_B(sk)m_h}{2k^2q}\left[
			\frac{k^2 I^+_1(\omega_{s,h}) +kq I^+_2(\omega_{s,h})+q^2I^+_3(\omega_{s,h})/4}{m^2_h}+I^+_1(\omega_{s,h})\right]\nonumber\\
			&&
			+g^2\int\frac{k\mathrm{d}k}{2\pi}~
			\frac{n_B(\xi_{h,{\bf k}})\xi}{2kq}\left[\frac{(k^2+q^2/4)J_1(\omega_h)+kqJ_2(\omega_h)}{m^2_h}+J_1(\omega_h)\right]\nonumber\\
			&&-\frac{g^2}{ m^2_h}\int\frac{kdk}{2\pi}~
			\frac{n_B(\xi_{h,{\bf k}})kq[J_1(\omega_h)-J_3(\omega_h)]}{8(\omega+i0^+-\xi_{h,k})^2},
		\end{eqnarray}
	\end{widetext}
	where 
 \begin{eqnarray}
     \omega_{s,h}=\frac{m_h}{kq}\left(\omega+sk-\frac{k^2+q^2}{2m_h}+\mu_h\right),
 \end{eqnarray}
 and
 \begin{eqnarray}
     \omega_{h}=\frac{\xi}{2kq}\left((\omega-\xi_{h,{\bf k}})^2-k^2-q^2 \right),
 \end{eqnarray}
and the functions $J_{1,2,3}(\omega_h)$ are similar to $J_{1,2,3}(\omega_f)$.

\section{Polarization functions $\Pi^{f,h}_{ab}$}\label{Appendix:Polarization}

In this appendix we  {will} now calculate $\Pi^{f,h}_{ab}({\bf q},\omega)$ using  perturbation theory. There are two  {types} of vacuum polarization for both  spinon and holon. 
\begin{eqnarray}
	\Pi_{fab}^1({\bf q},i\omega_n)&=& \frac{g^2}{4m_f^2(2\pi)^2\beta N}\sum_{\sigma,m}\int d^2p\nonumber\\
	&&{\cal G}_{f\sigma}({\bf p+q},i\omega_n+i\omega'_m)\nonumber\\
	&&{\cal G}_{f\sigma}(-{\bf p},-\omega')(2p_a+q_a)(-2p_b-q_b),~~\\
	\Pi_{hab}^1({\bf q},i\omega_n)&=& \frac{g^2}{4m_h^2(2\pi)^2\beta N}\sum_m\int  d^2p\nonumber\\
	&&{\cal G}_h({\bf p+q},i\omega_n+i\omega'_m)\nonumber\\
	&&{\cal G}_h(-{\bf p},-\omega')(2p_a+q_a)(-2p_b-q_b),\\
	\Pi_{fab}^2({\bf q},\omega)&=& \frac{-g^2}{2m_f(2\pi)^2\beta N}\sum_{\sigma,m}\int d^2p{\cal G}_{f\sigma}({\bf p},i\omega'_m)\delta_{ab},\nonumber\\
	&&\\
	\Pi_{hab}^2(q,\omega)&=& \frac{-g^2}{2m_h(2\pi)^3}\sum_m\int d^2p{\cal G}_h({\bf p},i\omega'_m)\delta_{ab}.
\end{eqnarray}

To the order of $O(g^2)$, all spinon and holon Green's functions are approximated by free ones. Then
\begin{eqnarray}
	\Pi_{fab}^1(q,i\omega_m)&=&\frac{g^2}{2m_f^2(2\pi)^2 N}\int d^2p
	\frac{n_F(\xi_{f,\bf p})-n_F(\xi_{f,\bf p+q})}{i\omega_m+\xi_{f,\bf p}-\xi_{f ,\bf p+q}}\nonumber\\
	&&(2p_a+q_a)(-2p_b-q_b), \label{Pif1}\\
	\Pi_{fab}^2(q,i\omega_m)&=&\frac{-g^2}{m_f(2\pi)^2N}\int d^2 p n_F(\xi_{f,\bf p})\delta_{ab},\\
	\Pi_{hab}^1(q,i\omega_m)&=&\frac{-g^2}{4m_h^2(2\pi)^2 N}\int d^2p
	\frac{n_B(\xi_{h,\bf p})-n_B(\xi_{h,\bf p+q})}{i\omega_m+\xi_{h,\bf p}-\xi_{h,\bf p+q}}\nonumber\\
	&&(2p_a+q_a)(-2p_b-q_b),\\
	\Pi_{hab}^2(q,i\omega_m)&=&\frac{g^2}{2m_h(2\pi)^2N}\int d^2 p n_B(\xi_{h,\bf p})\delta_{ab}.
\end{eqnarray}

	\bibliography{BRST}

\end{document}